\documentclass[11pt,a4paper]{article}
\usepackage{jheppub}

\usepackage{amsmath}
\usepackage{bbm}
\usepackage{float}
\usepackage{graphicx}	
\usepackage{hyperref}
\usepackage{mathtools}

\usepackage{multirow}
\usepackage{rotating}
\usepackage{subcaption}

\def\thetitle{Compact Perturbative Expressions for  Neutrino Oscillations in Matter: II}
\title{\thetitle}
\hypersetup{pdftitle={\thetitle}}

\author[a]{Peter B.~Denton,}
\author[b]{Hisakazu Minakata,}
\author[c]{Stephen J.~Parke}

\affiliation[a]{Niels Bohr International Academy, Niels Bohr Institute, University of Copenhagen,\\
         Blegdamsvej 17, 2100, 
         Copenhagen, Denmark}
\affiliation{
$^b$Instituto F\'{\i}sica Te\'{o}rica, UAM/CSIC, Calle Nicola's Cabrera 13-15, Cantoblanco E-28049 Madrid, Spain. \&
 Research Center for Cosmic Neutrinos, Institute for Cosmic Ray Research, University of Tokyo, Kashiwa, Chiba 277-8582, Japan 
}

\affiliation[c]{Theoretical Physics Department, Fermi National Accelerator Laboratory, P.~O.~Box 500, Batavia, IL 60510, USA}

\emailAdd{peterbd1@gmail.com}
\emailAdd{hisakazu.minakata@gmail.com}
\emailAdd{parke@fnal.gov}

\preprint{FERMILAB-PUB-18-014-T}

\abstract{In this paper we rewrite the neutrino mixing angles and mass squared differences in matter given in our original paper, \cite{Denton:2016wmg}, in a notation that is more conventional for the reader.
  Replacing the usual neutrino mixing angles and mass squared differences in the expressions for the vacuum oscillation probabilities with these matter mixing angles and mass squared differences gives  an excellent approximation to the oscillation probabilities in matter. Comparisons for T2K, NOvA, T2HKK and DUNE are also given for neutrinos and anti-neutrinos, disappearance and appearance channels, normal ordering and inverted ordering.}


\numberwithin{equation}{subsection}

\begin{document}

\maketitle

\section{Introduction}
In this paper we rewrite the neutrino mixing angles and mass squared differences in matter given in our original paper, \cite{Denton:2016wmg}, in a notation that is more conventional for the reader.
  Replacing the usual neutrino mixing angles and mass squared differences in the expressions for the vacuum oscillation probabilities with these matter mixing angles and mass squared differences gives  an excellent approximation to the oscillation probabilities in matter.  Higher orders are also easily calculated and provide several orders of magnitude improvement per order.

In Section 2, we give the approximation to the mixing angles and mass squared difference in matter and discuss how to use these to calculated the oscillation probabilities in matter both at 0th order and 1st order. We also give expansions of the mixing angles and mass squared differences in matter in powers of $(a/\Delta m^2)$.  In Section 3, we make a detailed comparison between the exact and the approximate oscillation probabilities in matter for the T2K, NOvA, T2HKK and DUNE experiments.  Section 4 is the Summary.

\newpage

\section{Mixing Angles and Mass Differences in Matter}
\subsection{Zeroth Order}

In this section, a simple and accurate way to evaluate oscillation probabilities, recently shown  in \cite{Denton:2016wmg}, is given.\footnote{In this note $\phi$, $\psi$ and $\Delta \lambda_{jk}$ of \cite{Denton:2016wmg}, are replaced with the more traditional notation $\widetilde{\theta}_{13}$  and $\widetilde{\theta}_{12}$ and $\Delta \, \widetilde{m^2}_{jk}$ respectively.} Details as to the why's and how's of this method are contained in that paper.\\

The mixing angles in matter, which we denote by a $\widetilde{\theta}_{13}$  and $\widetilde{\theta}_{12}$ here, can also be calculated in the following way, using 
$ \Delta m^2_{ee} \equiv \cos^2 \theta_{12} \Delta m^2_{31} + \sin^2 \theta_{12} \Delta m^2_{32}$,  as follows\footnote{Vacuum values to be used in calculating $\Delta m^2_{ee}$.}, see \cite{Parke:2018brr}:
\begin{eqnarray}
\cos 2 \widetilde{\theta}_{13} & = & \frac{ (\cos 2\theta_{13} -a/\Delta m^2_{ee}) } 
{  \sqrt{(\cos 2\theta_{13}-a/\Delta m^2_{ee})^2 +  \sin^22\theta_{13}  ~ }}, 
 \label{eq:th13}  
 \end{eqnarray}
 where  $~~a  \equiv   2 \sqrt{2} G_F N_e E_\nu~~$  is the standard matter potential, and
\begin{eqnarray}
 \cos 2 \widetilde{\theta}_{12} & = &  \frac{ ( \cos 2\theta_{12} 
 - a^{\,\prime}  /\Delta m^2_{21} ) } {  \sqrt{(\cos 2\theta_{12} 
 -a^{\,\prime} /\Delta m^2_{21})^2 ~+~
  \sin^2 2 \theta_{12} \cos^2( \widetilde{\theta}_{13}-\theta_{13})~~}  }, \label{eq:th12} 
  \end{eqnarray}
where 
$~~a^{\,\prime}    \equiv   a \, \cos^2 \widetilde{\theta}_{13} +\Delta m^2_{ee} \sin^2  ( \widetilde{\theta}_{13}-\theta_{13} )~~$
is the $\theta_{13}$-modified matter potential for the 1-2 sector.
In these two flavor rotations, both $\widetilde{\theta}_{13}$ and  $\widetilde{\theta}_{12}$ are in range $[0,\pi/2]$.\\

$\theta_{23}$ and $\delta$ are unchanged in matter for this approximation.\\

The neutrino mass squared differences in matter, i.e. the $\Delta m^2_{jk}$ in matter, which we denote by $\Delta \, \widetilde{m^2}_{jk}$, are given by 
   \begin{eqnarray}
  \Delta\, \widetilde{m^2}_{21}  & = & \Delta m^2_{21} \, \sqrt{(\cos 2\theta_{12} 
 - a^{\,\prime} /\Delta m^2_{21})^2 ~+~
  \sin^2 2 \theta_{12} \cos^2(\widetilde{\theta}_{13}-\theta_{13})~~} , \nonumber   \\[1mm]
   \Delta\,  \widetilde{m^2}_{31}   &=& \Delta m^2_{31} + ( \, a-\frac{3}{2}a^{\, \prime} \, ) +\frac{1}{2}\,\left( 
  \, \Delta \widetilde{m^2}_{21}  -\Delta m^2_{21}   ~\right) , \label{eq:dmsqa}   \\[1mm]
   \Delta\,  \widetilde{m^2}_{32}  & = &  \Delta \,  \widetilde{m^2}_{31} -\Delta\,   \widetilde{m^2}_{21}
   \nonumber
   %
   %
 \end{eqnarray}
Note that the same square root\footnote{If $a=0$, then $ \widetilde{\theta}_{13}=  \theta_{13}$ and since $a^\prime=0$ then $ \widetilde{\theta}_{12}=  \theta_{12}$ and both $\sqrt{\cdots}=1$, also $\Delta \, \widetilde{m^2}_{jk}=\Delta m^2_{jk}$ for all $(j,k)$ as required. The identity $s^2_\theta=(1-\cos 2 \theta)/2$ is useful for calculating both $s_\theta$ and $c_\theta$. } 
appears  in both $ \Delta \, \widetilde{m^2}_{21}$ and $\sin^2 \widetilde{\theta}_{12}$.
  To see that the $\Delta\,  \widetilde{m^2}_{31}$ and  $\Delta\,  \widetilde{m^2}_{32}$ have the right asymptotic forms, use 
 the fact that $ (\Delta \, \widetilde{m^2}_{21}  -\Delta m^2_{21}) = |a^{\, \prime} | +{\cal O}(\Delta m^2_{21})$, for $|a| \gg \Delta m^2_{21}$.
 
\begin{figure}[t]
\begin{center}
\vspace*{-1.0cm}
     \includegraphics[width=.48\textwidth]{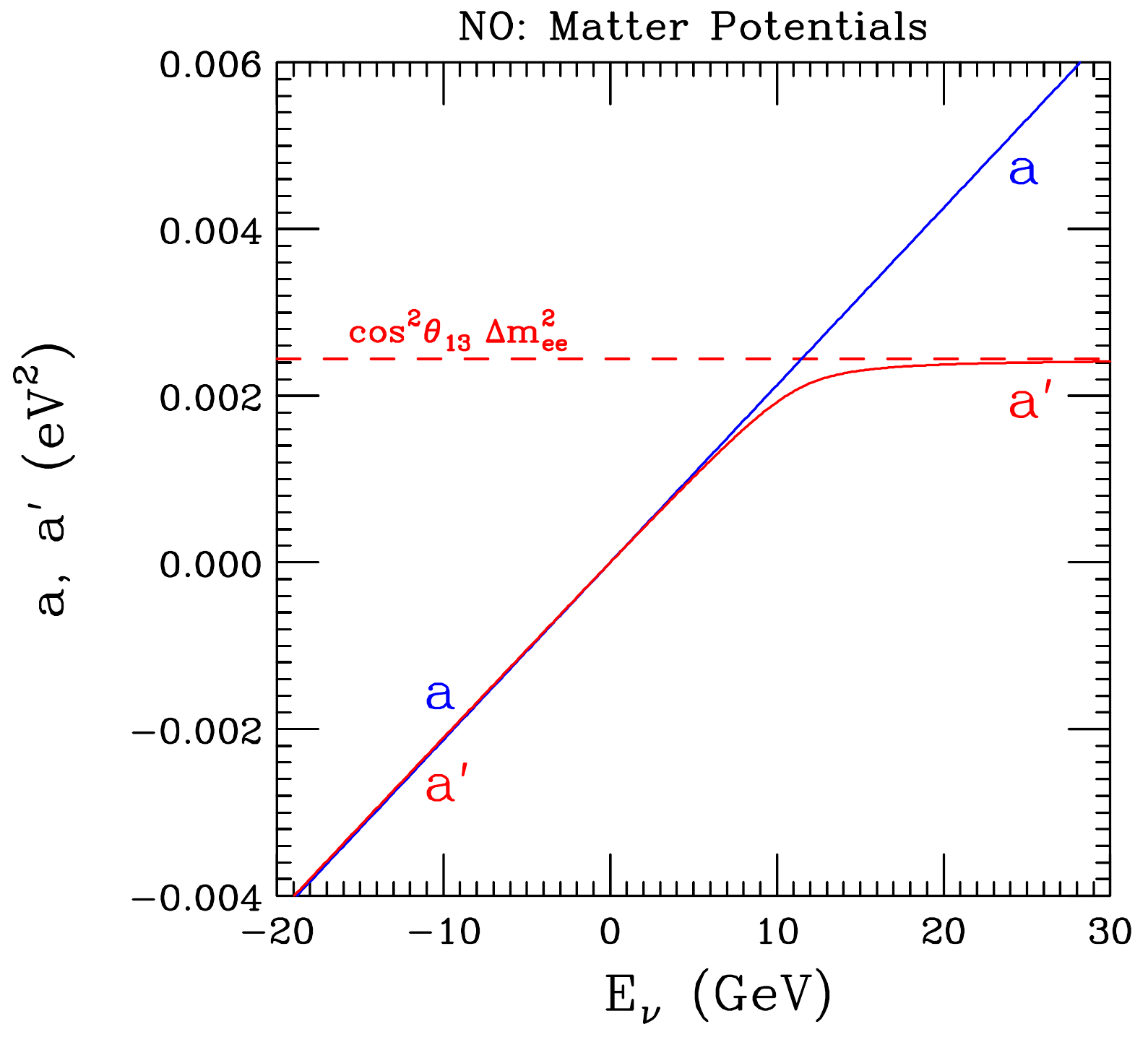}
       \includegraphics[width=.45\textwidth]{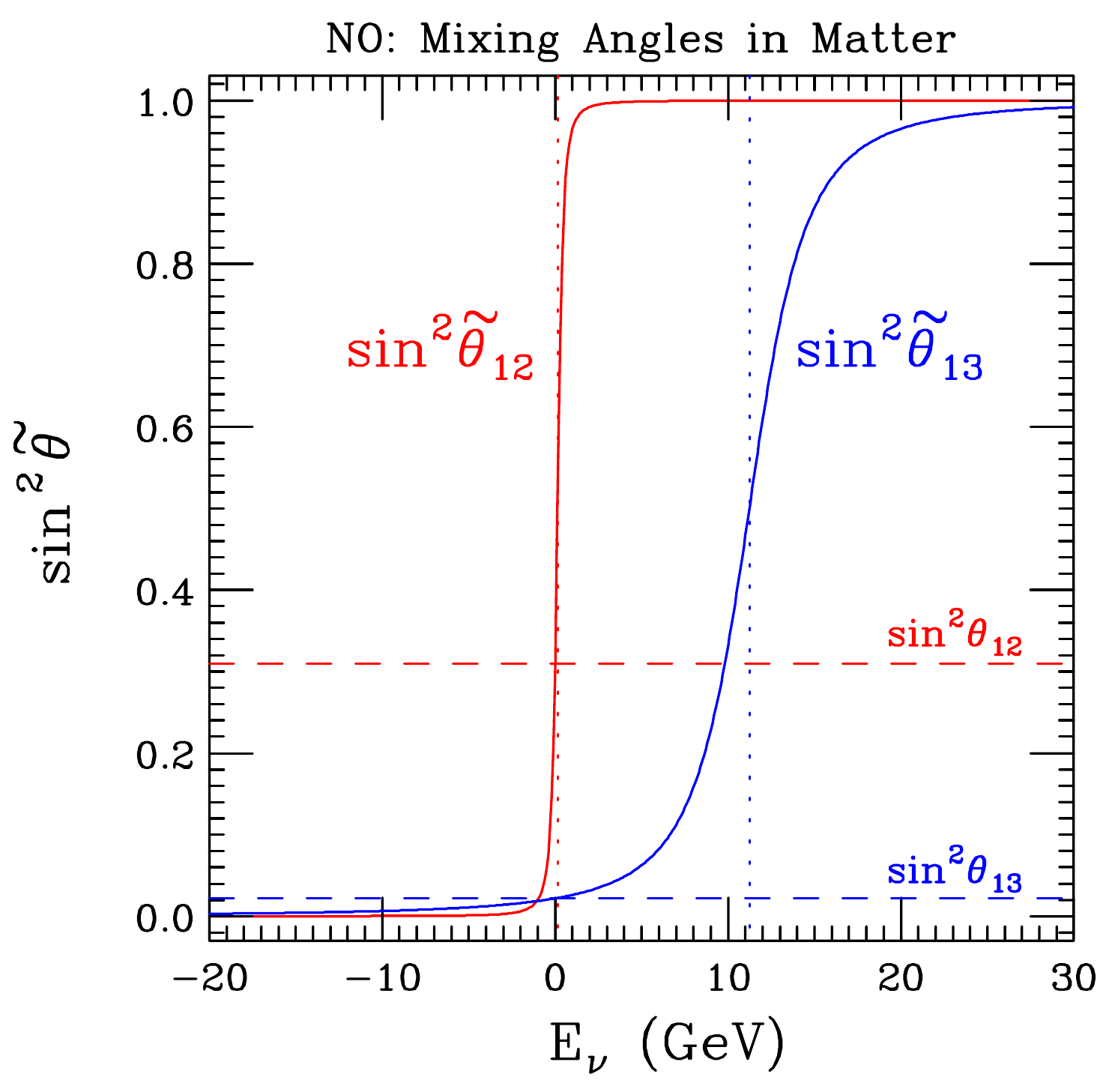}\\
      \includegraphics[width=.48\textwidth]{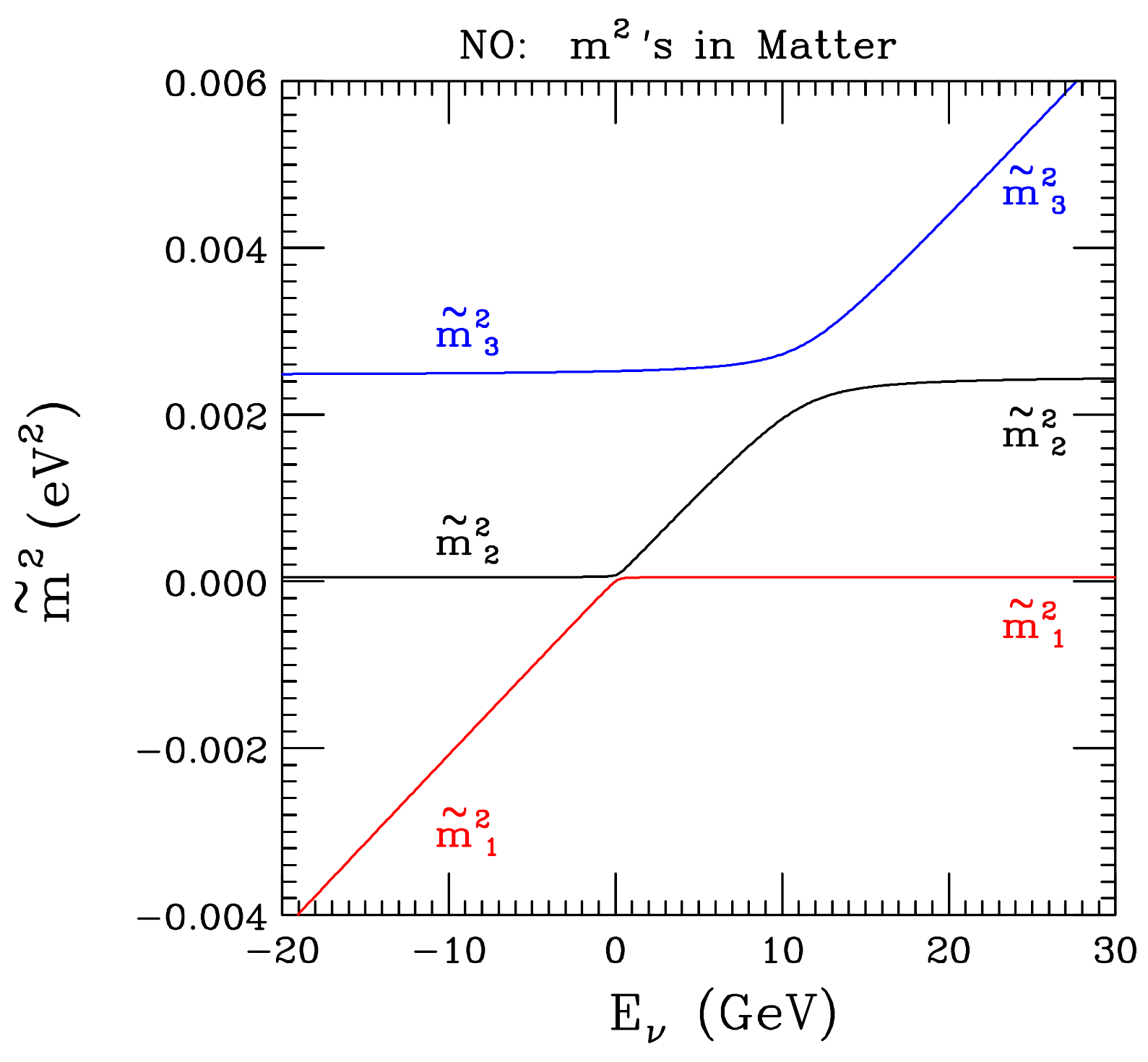}
       \includegraphics[width=.48\textwidth]{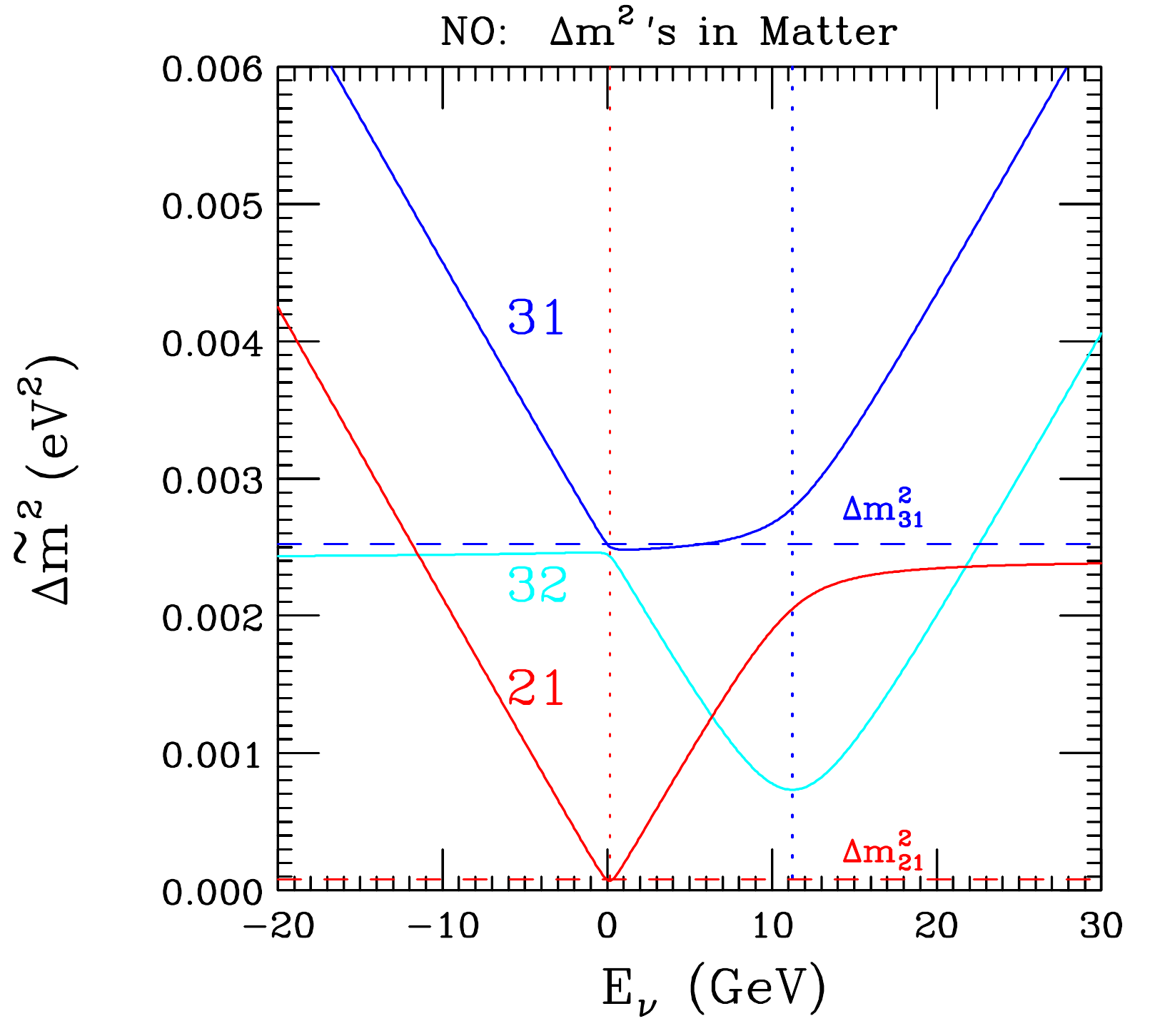}
  \caption{In the normal ordering (NO): Top left, the matter potentials, $a$ and $a^{\, \prime}$, top right, sine squared of mixing angles in matter, $ \sin^2 \widetilde{\theta}_{jk}$, bottom left,  the mass squared eigenvalues in matter,  $\widetilde{m^2}_{j}$, and bottom right, the mass squared differences in matter, $ \Delta \, \widetilde{m^2}_{jk}$. $E_\nu \geq 0  ~(E_\nu \leq 0) $ is for neutrinos (anti-neutrinos). $E_\nu=0$ is the vacuum values for both neutrinos and anti-neutrinos.}
     \label{fig:NO}
          \end{center}

     \end{figure}

\begin{figure}[h]
\begin{center}
\vspace*{-1.5cm}
     \includegraphics[width=.47\textwidth]{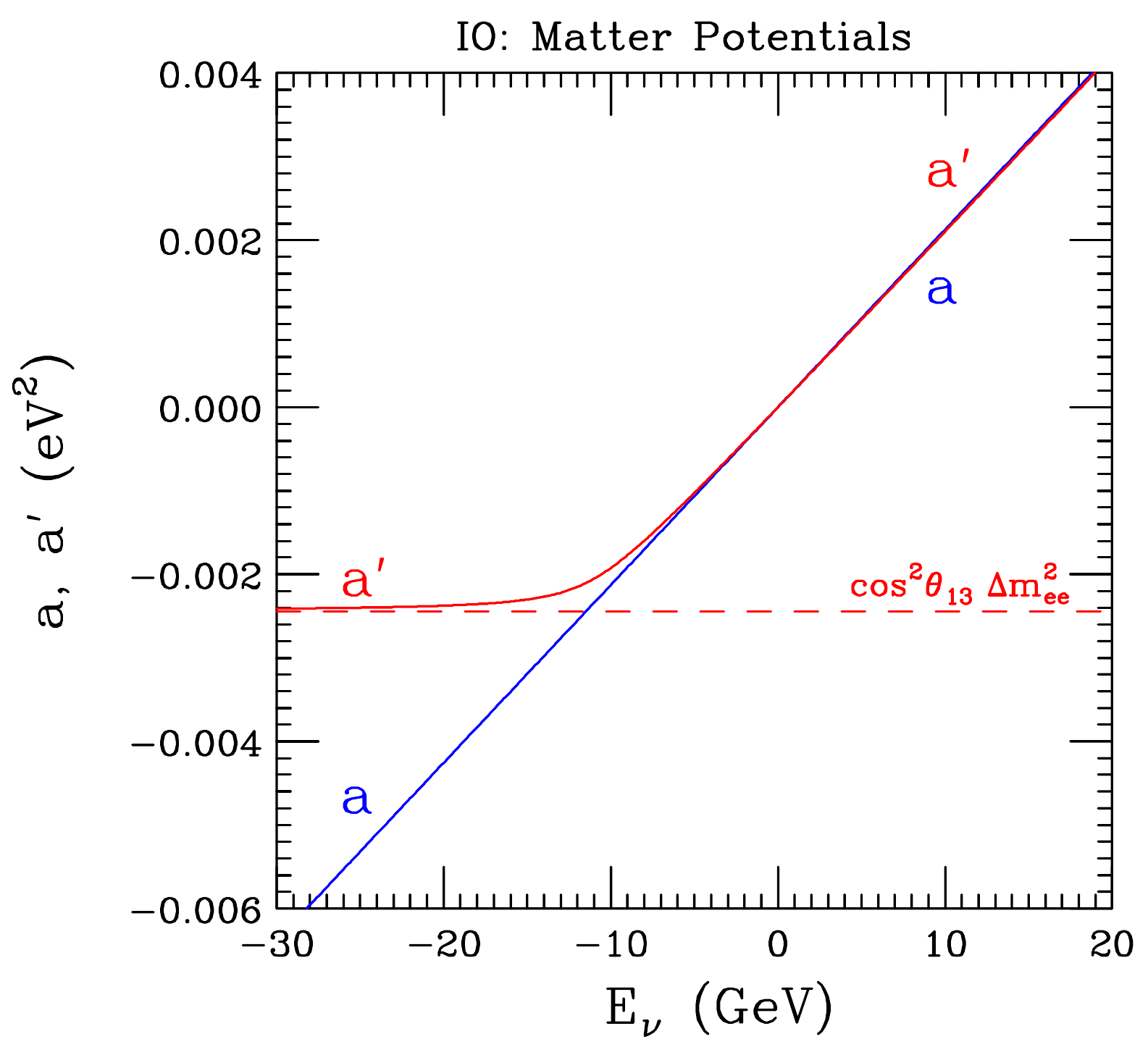}
       \includegraphics[width=.44\textwidth]{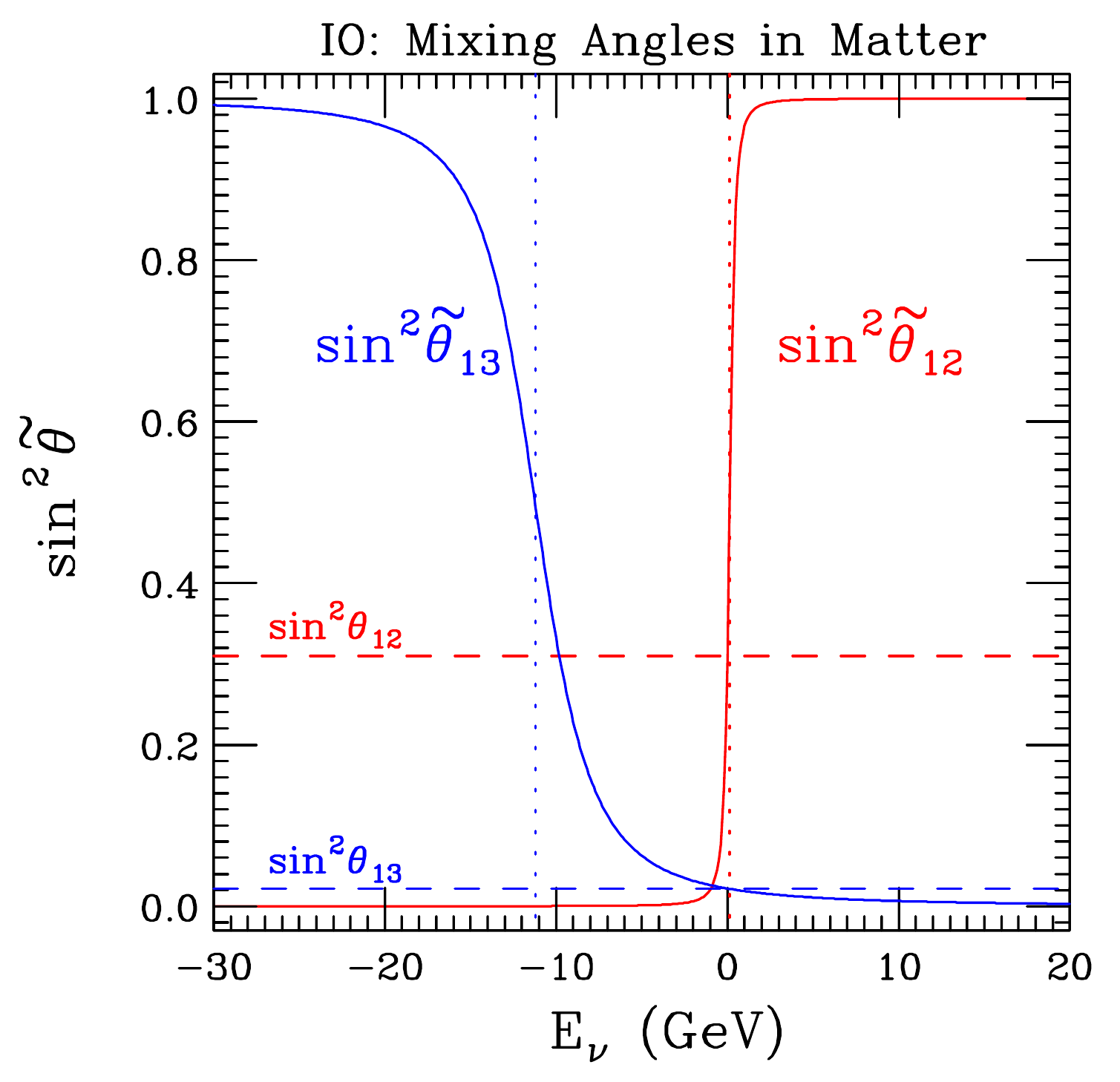}\\
      \includegraphics[width=.47\textwidth]{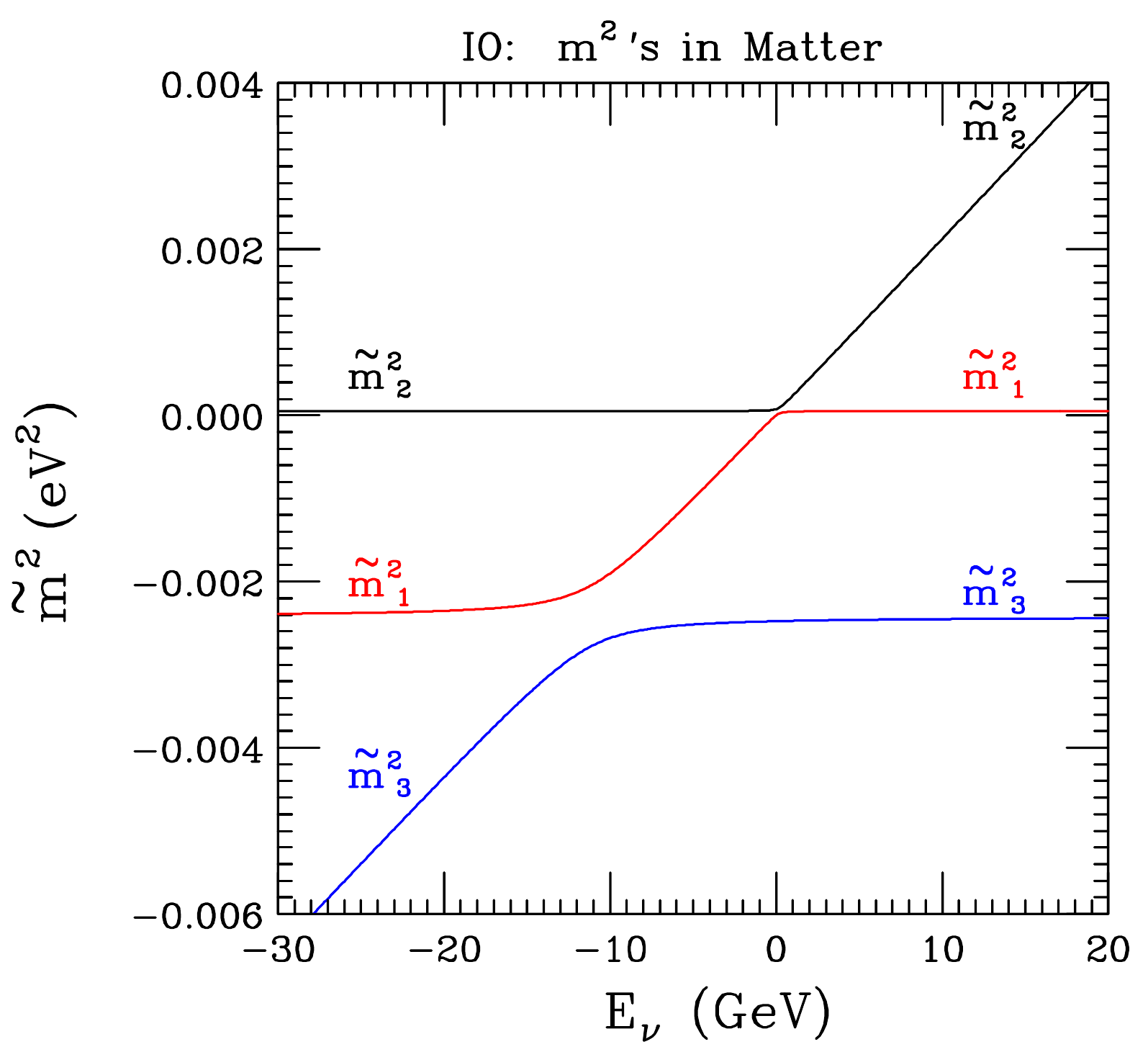}
       \includegraphics[width=.47\textwidth]{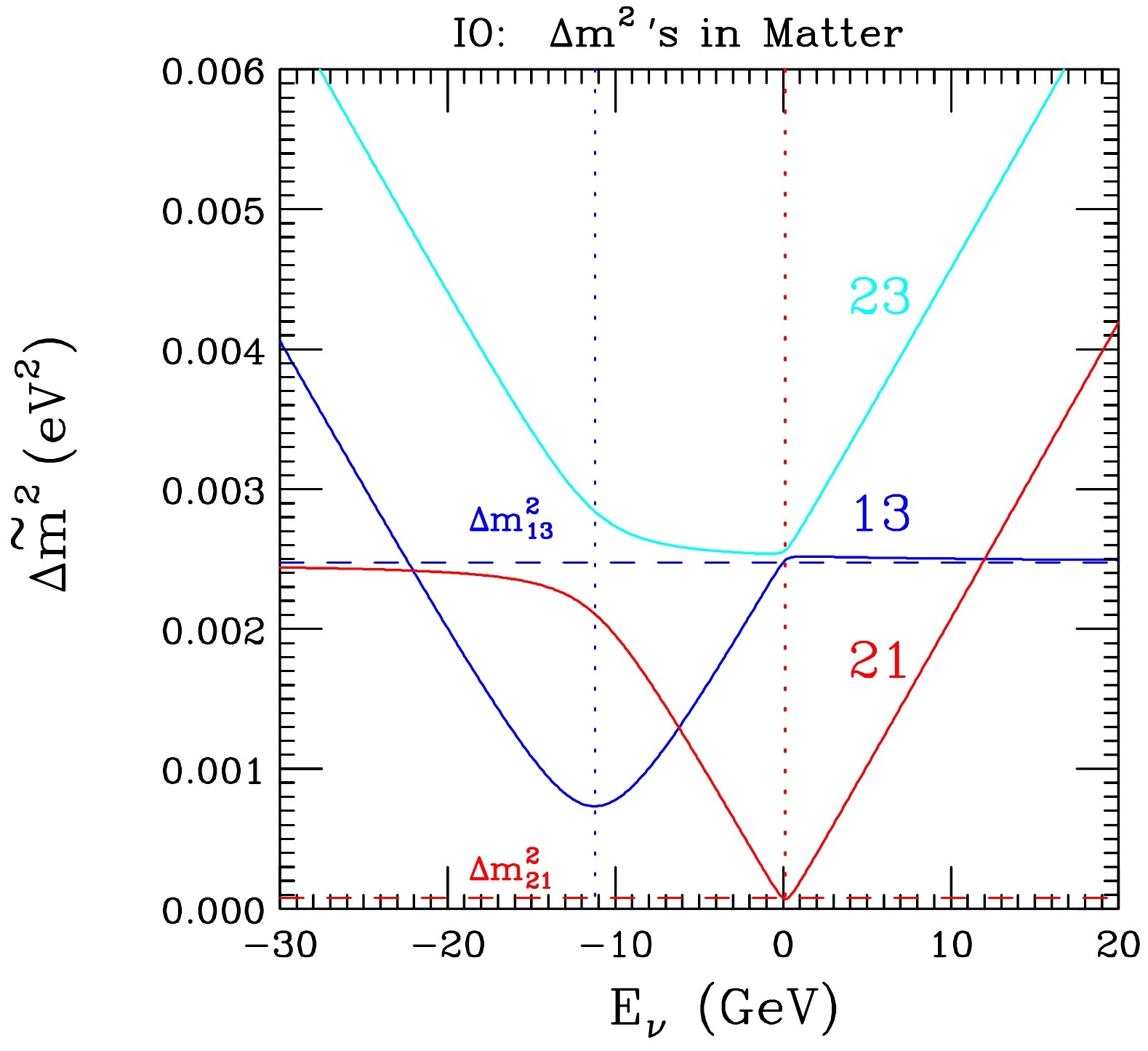}
  \caption{In the inverted ordering (IO):  Top left, the matter potentials, $a$ and $a^{\, \prime}$, top right, sine squared of mixing angles in matter, $ \sin^2 \widetilde{\theta}_{jk}$, bottom left,  the mass squared eigenvalues in matter,  $\widetilde{m^2}_{j}$, and bottom right, the mass squared differences in matter, $ \Delta \, \widetilde{m^2}_{jk}$. $E_\nu \geq 0  ~(E_\nu \leq 0) $ is for neutrinos (anti-neutrinos). $E_\nu=0$ is the vacuum values for both neutrinos and anti-neutrinos.}
      \label{fig:IO}
          \end{center}

     \end{figure}

In Fig. \ref{fig:NO} and Fig. \ref{fig:IO} the values of a, a$^{\,\prime}$, $\sin^2 \widetilde{\theta}_{13}$, $\sin^2 \widetilde{\theta}_{12}$,
$\widetilde{m^2}_j$ and $\Delta \widetilde{m^2}_{jk}$ as a function of the neutrino energy for a density of 3.0 g.cm$^{-3}$. \\


To calculate the oscillation probabilities, to 0th order,  use the above $ \Delta \, \widetilde{m^2}_{jk} $ instead of $\Delta m^2_{jk}$ and replace the vacuum MNS matrix as follows
\begin{eqnarray}
U^0_{MNS} \equiv U_{23}(\theta_{23})\,U_{13}(\theta_{13},\delta)\,U_{12}(\theta_{12})  
& \Rightarrow  & U^M_{MNS} \equiv U_{23}(\theta_{23})\,U_{13}( \ \widetilde{\theta}_{13},\delta)\,U_{12}(  \widetilde{\theta}_{12}).
\nonumber
\end{eqnarray}
That is, replace
\begin{eqnarray}
\Delta m^2_{jk}  & \rightarrow &  \Delta \, \widetilde{m^2}_{jk}  \nonumber \\
\theta_{13}  & \rightarrow &  \widetilde{\theta}_{13} \nonumber \\
\theta_{12}  & \rightarrow &  \widetilde{\theta}_{12},
\end{eqnarray}
$\theta_{23}$ and $\delta$ remain unchanged, it is that simple. We call this the 0th order DMP approximation.

These expressions are valid for both NO, $\Delta m^2_{ee}>0$ and IO,
$\Delta m^2_{ee}<0$.  For anti-neutrinos, just change the sign of $a$ and $\delta$.
Our expansion parameter is 
\begin{eqnarray}
\left| \, \sin( \widetilde{\theta}_{13}-\theta_{13}) ~\sin \theta_{12} \cos \theta_{12} ~\frac{\Delta m^2_{21}}{\Delta m^2_{ee}} \, \right| \leq 0.015,
\end{eqnarray}
which is small and vanishes in vacuum, so that our perturbation theory reproduces the vacuum oscillation probabilities exactly.\\

If $P_{\nu_\alpha \rightarrow \nu_\beta}( \Delta m^2_{31}, \Delta m^2_{21}, \theta_{13}, \theta_{12},\theta_{23},\delta)$ is the oscillation probability in vacuum
 then  \\[1mm]
$P_{\nu_\alpha \rightarrow \nu_\beta}( \Delta \, \widetilde{m^2}_{31}, \Delta \, \widetilde{m^2}_{21}, \widetilde{\theta}_{13}, \widetilde{\theta}_{12},\theta_{23},\delta)$ is the oscillation probability in matter, i.e. use the same function but replace the mass squared differences and mixing angles with the matter values given in eq. \ref{eq:th13} - \ref{eq:dmsqa}. The resulting oscillation probabilities are identical to the zeroth order approximation given in  Denton, Minakata and Parke, \cite{Denton:2016wmg}.

\subsection{Higher Orders}
If the 0th order is not accurate enough, going to 1st order is simple and gives another two orders of magnitude in accuracy.  
First the $\Delta \, \widetilde{m^2}_{jk} $ remain unchanged but the mixing matrix is modified by
\begin{eqnarray}
U^M_{MNS} & \Rightarrow & V \equiv U^M_{MNS} (1+W_1),
\end{eqnarray}
where the matrix $W_1$ is given by
\begin{eqnarray}
W_1 & =  &\sin( \widetilde{\theta}_{13}-\theta_{13}) ~s_{12} c_{12} ~\Delta m^2_{21} \nonumber \\[3mm]
& & \left( \begin{array}{ccc}
 0 & 0 & -\widetilde{s}_{12} \,e^{-i\delta} /  \Delta \, \widetilde{m^2}_{31} \\
 0 &0 & +\widetilde{c}_{12}\,e^{-i\delta} / \ \Delta \, \widetilde{m^2}_{32} \\
 +\widetilde{s}_{12}\, e^{+i\delta}  /  \Delta \, \widetilde{m^2}_{31} \quad  & -\widetilde{c}_{12}\, e^{+i\delta} /  \Delta \, \widetilde{m^2}_{32} \quad &  0
 \end{array} \right). 
 \end{eqnarray}
 where $\widetilde{s}_{12}=\sin  \widetilde{\theta}_{12}$ and $\widetilde{c}_{12}=\cos  \widetilde{\theta}_{12}$ etc.
 The $\Delta \, \widetilde{m^2}_{jk} $ and the $V$-mixing matrix can be used to calculate the oscillation probabilities and improves the accuracy by two orders of magnitude.   We call this the 1st order DMP approximation. The next highest order, 2nd order, is also discussed in \cite{Denton:2016wmg}.

\subsection{Expansions in $a/\Delta m^2$}

If $|a| \ll | \Delta m^2_{ee} |$,  that is when $E_\nu \ll 11 \left(\, \rho\, /\,3 ~{\rm g\,cm^{-3}} \right)$ GeV,  we have, \begin{eqnarray}
\sin^2 \widetilde{\theta}_{13}  & \approx & s^2_{13} \biggr[1+ 2  c^2_{13} (a/\Delta m^2_{ee}) +3(c^2_{13}-s^2_{13})c^2_{13} (a/\Delta m^2_{ee})^2 +  {\cal O}(a/\Delta m^2_{ee})^3) \biggr] \nonumber  \\
 \sin^2( \widetilde{\theta}_{13}-\theta_{13}) & \approx &  s^2_{13}c^2_{13} (a/\Delta m^2_{ee})^2\biggr[1+ 2 (c^2_{13}-s^2_{13}) (a/\Delta m^2_{ee}) + {\cal O}(a/\Delta m^2_{ee})^2 \biggr] \\
a^{\,\prime} & \approx & a c^2_{13} \biggr[1- s^2_{13} (a/\Delta m^2_{ee}) -  s^2_{13}(c^2_{13}-s^2_{13}) (a/\Delta m^2_{ee})^2  +   {\cal O}(a/\Delta m^2_{ee})^3\biggr] \nonumber
\end{eqnarray}
up to ${\cal O}(a/\Delta m^2_{ee})^2$.  The expansion for $a^{\,\prime}$ can be used to calculate $ \Delta\,  \widetilde{m^2}_{31}  $ as follows,
\begin{eqnarray}
 \Delta\,  \widetilde{m^2}_{31}   &=&\left\{\begin{array}{ll}
  \Delta m^2_{31} + ~( \, a- a^{\, \prime} \, ) +\frac{1}{2}\,\left[ 
  \, \Delta \widetilde{m^2}_{21}  -\Delta m^2_{21} - a^{\, \prime}    ~\right], & ~~a,  a^{\, \prime}>0 \\[2mm]
\Delta m^2_{31} + ( \, a- 2a^{\, \prime} \, ) +\frac{1}{2}\,\left[ 
  \, \Delta \widetilde{m^2}_{21}  -\Delta m^2_{21} + a^{\, \prime}    ~\right], & ~~a,  a^{\, \prime}<0 
  \end{array}
  \right.
  \label{eq:dmsqa_rewrite}   
  \end{eqnarray}
where the quantities in $[\cdots]$ is of ${\cal O}(\Delta m^2_{21})$ for all values of $E_\nu$.

 As can be seen from Fig. \ref{fig:NO} and Fig. \ref{fig:IO}, both $  \Delta\,  \widetilde{m^2}_{21}  $ and $\sin^2 \widetilde{\theta}_{12}$ make rapid changes in +150 MeV region.   Well away from this region, when  $|a| \gg  | \Delta m^2_{21} |$, that is $E_\nu \gg 150  \left( \,\rho\, /\,3 ~{\rm g\,cm^{-3}} \right)$ MeV, we can write
\begin{eqnarray}
   \Delta\,  \widetilde{m^2}_{21}   & \approx &    | \,  a^{\, \prime} -  \Delta m^2_{21}\cos 2 \theta_{12}\,| \,,  
\end{eqnarray}
for $ | \,  ~a^{\, \prime} -  \Delta m^2_{21}\cos 2 \theta_{12}\,| \gg \Delta m^2_{21}$.
This can be used to obtain the asymptotic values for neutrino mass squareds in matter, which agree with the values given in  \cite{Denton:2016wmg}.

\section{Oscillation Probabilities}
\subsection{Comparisons}

Neutrino parameters relevant for oscillations:
\begin{eqnarray}
\Delta m^2_{ee} = \pm ~2.5 \times 10^{-3} ~{\rm eV}^2,  &  &\Delta m^2_{21} = +~7.5 \times 10^{-5} ~{\rm eV}^2 \nonumber \\
\sin^2 \theta_{12} =0.31,  & & \sin^2 \theta_{23} = 0.43 \nonumber  \\
\sin^2\theta_{13} = 0.022, & &\delta = -72^\circ  = -2 \pi/ 5
\end{eqnarray}
Where $\Delta m^2_{ee}>0$ gives a normal ordering (NO) neutrino spectrum and $\Delta m^2_{ee}<0$ for inverted ordering (IO). Note we have avoided the special points: $\theta_{23} =\pi/4$ as well as $\delta = 0, \pm \pi/2, \pi$, so as not to overestimate the precision.

\begin{figure}[t]
\begin{center}
     \includegraphics[width=.3\textwidth]{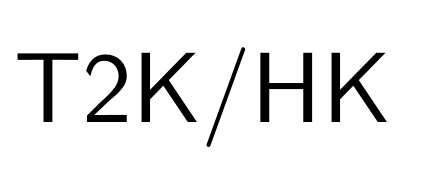}\\
     \includegraphics[width=.49\textwidth]{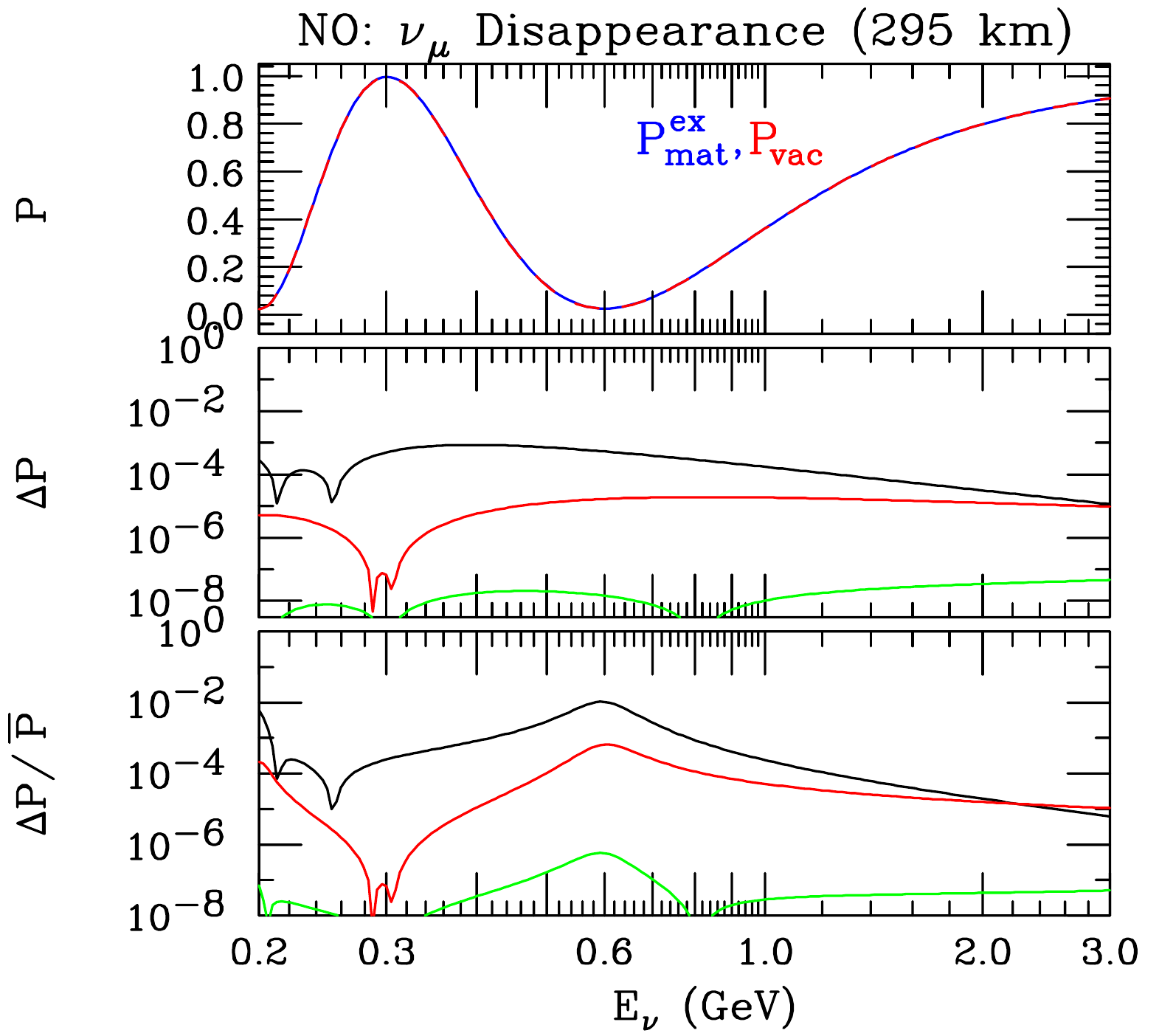}
     \includegraphics[width=.49\textwidth]{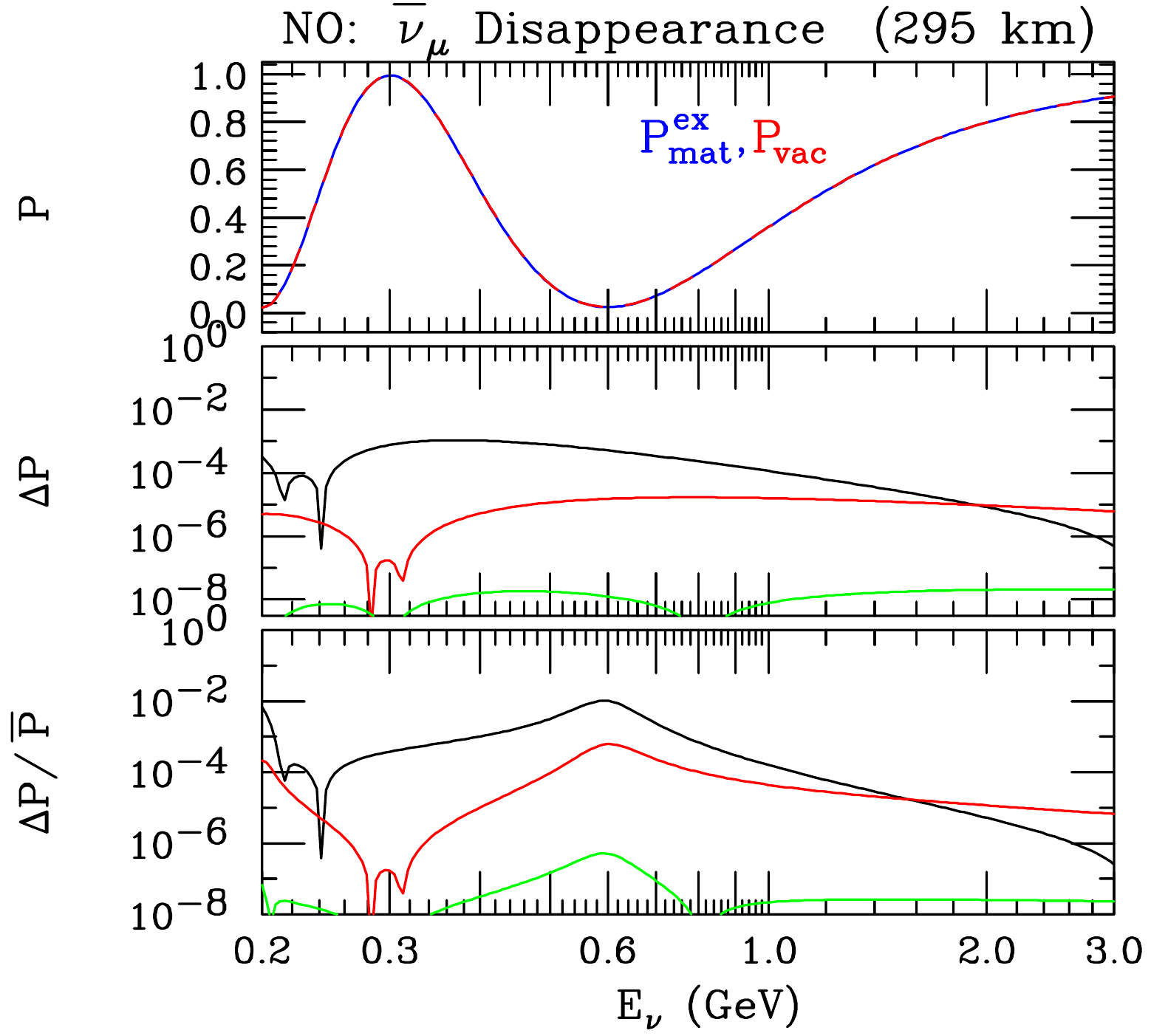}\\[5mm]
     \includegraphics[width=.49\textwidth]{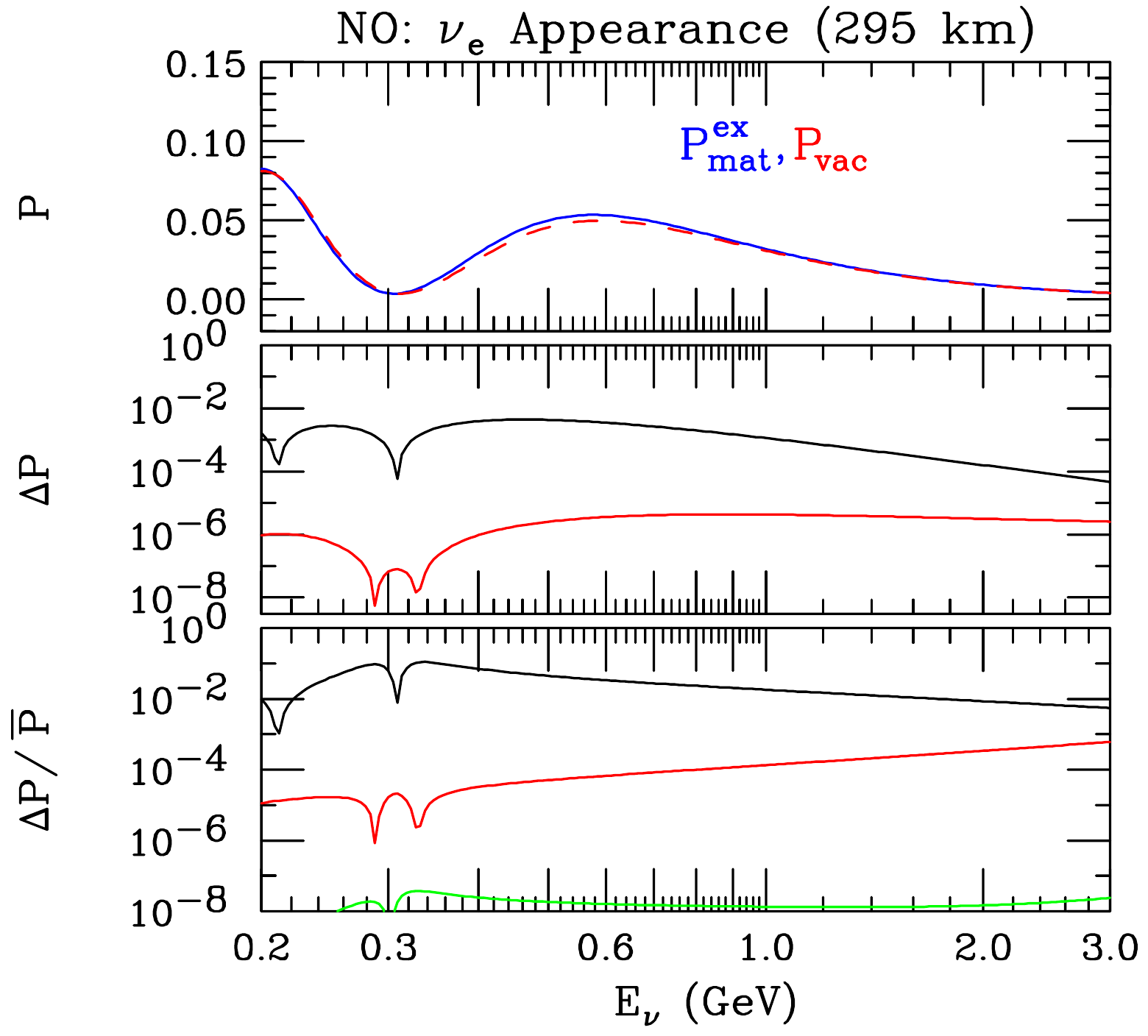}
     \includegraphics[width=.49\textwidth]{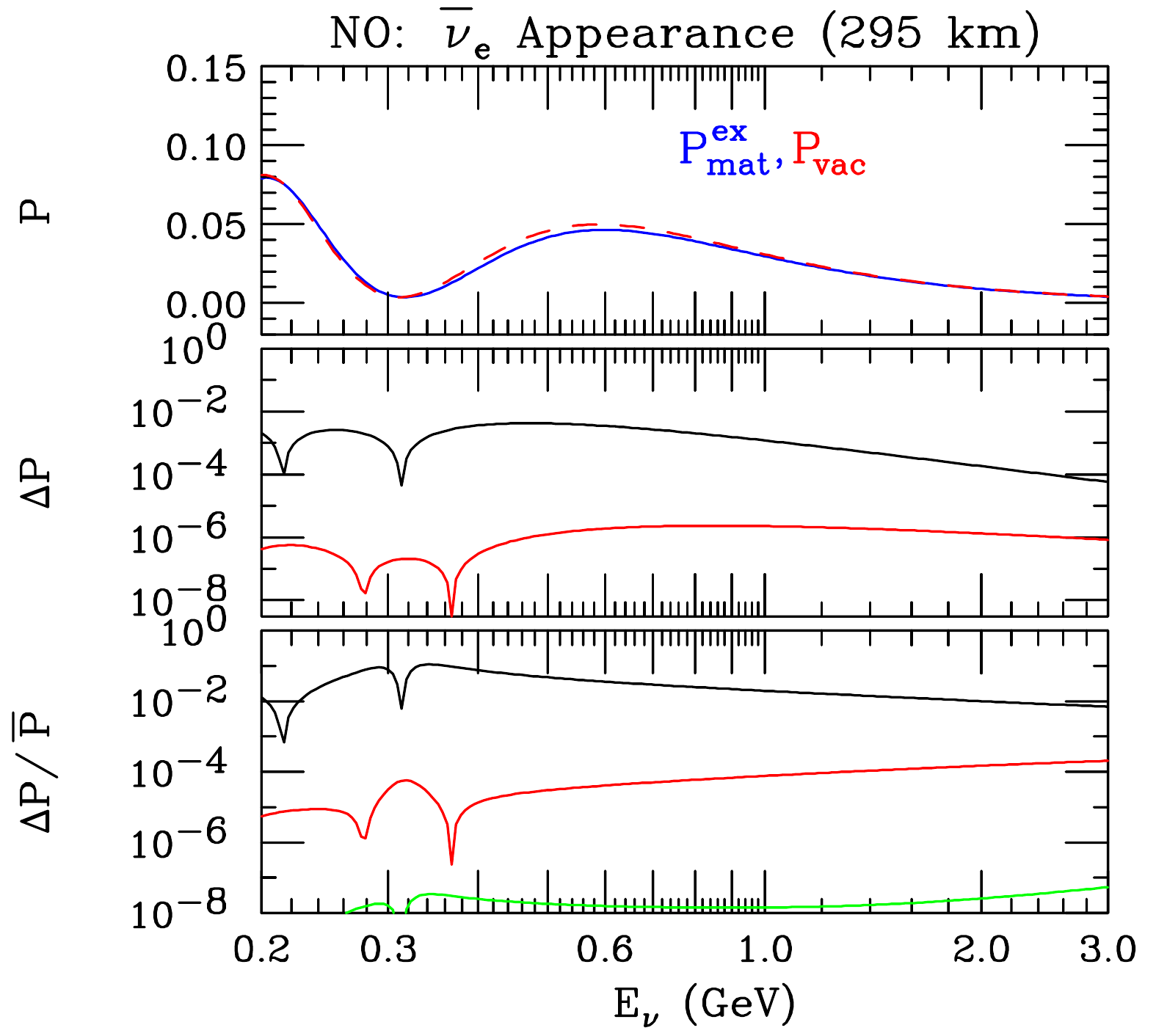}\\[5mm]
             \includegraphics[width=.95\textwidth]{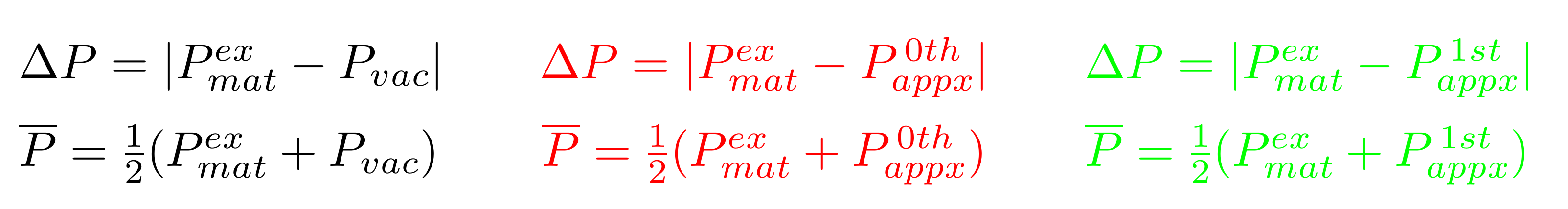}\\
  \caption{T2K, for normal ordering (NO): Top Left figure is $\nu_\mu$ disappearance, Top Right figure is $\bar{\nu}_\mu$ disappearance, Bottom Left figure is $\nu_\mu \rightarrow \nu_e$ appearance, and Bottom Right is  $\bar{\nu}_\mu \rightarrow \bar{\nu}_e$ appearance.
  In each figure, the top panel is exact oscillation probability in matter , $P^{ex}_{mat}$,  from \cite{Zaglauer:1988gz}, and in vacuum, $P_{vac}$. The Middle panel is difference between exact oscillation probabilities in matter and vacuum (black), and the difference between exact and 0th (red) and exact and 1st (green) approximations to the matter probabilities  using the DMP scheme, \cite{Denton:2016wmg}.  Bottom panel is similar to middle panel but plotting the fractional differences, $\Delta P/\overline{P}$. The density use is 2.3 g.cm$^{-3}$.
  }
     \label{fig:T2K_NO}
          \end{center}

     \end{figure}

\begin{figure}[t]
\begin{center}
     \includegraphics[width=.3\textwidth]{T2K.pdf}\\
     \includegraphics[width=.49\textwidth]{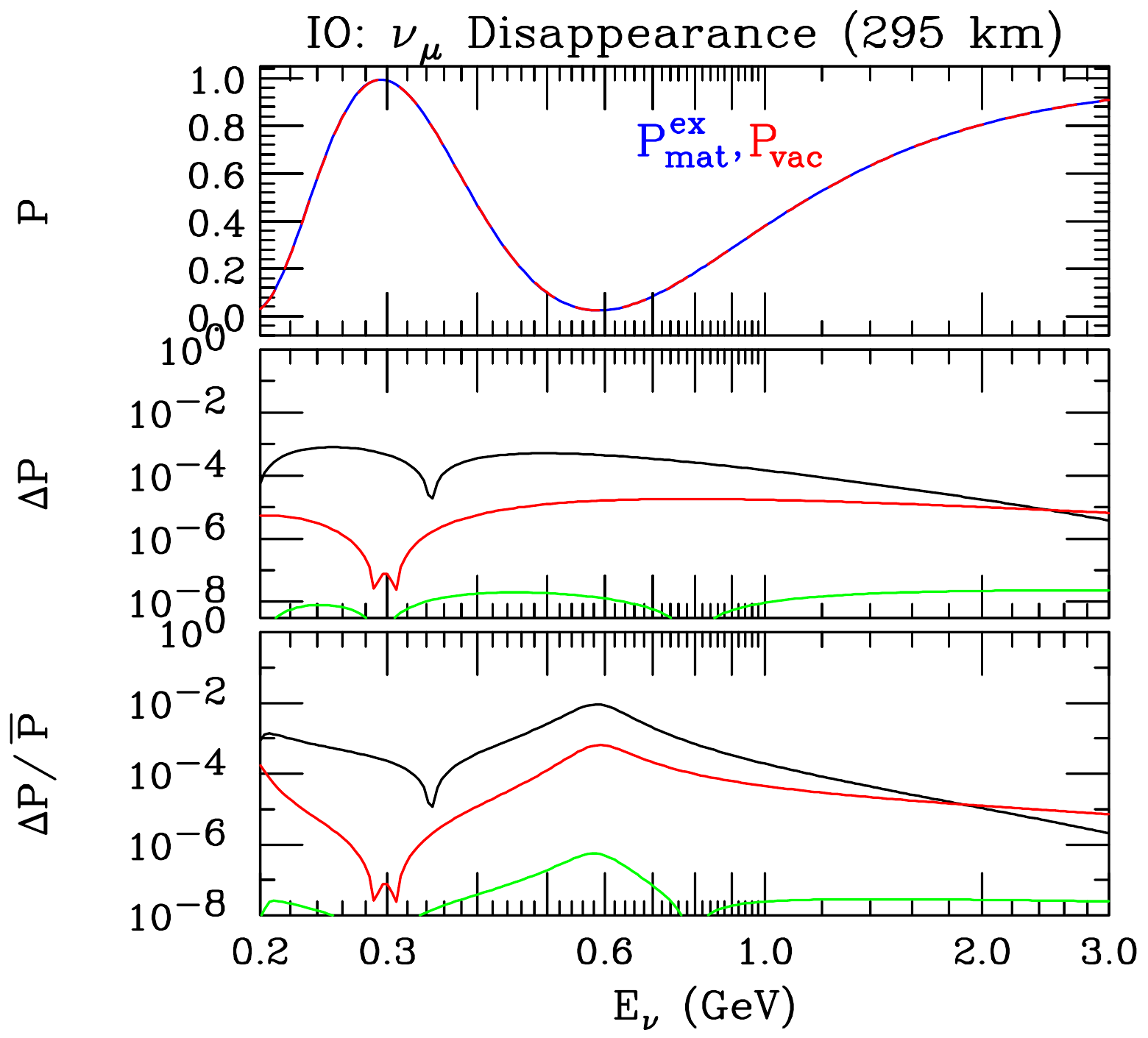}
     \includegraphics[width=.49\textwidth]{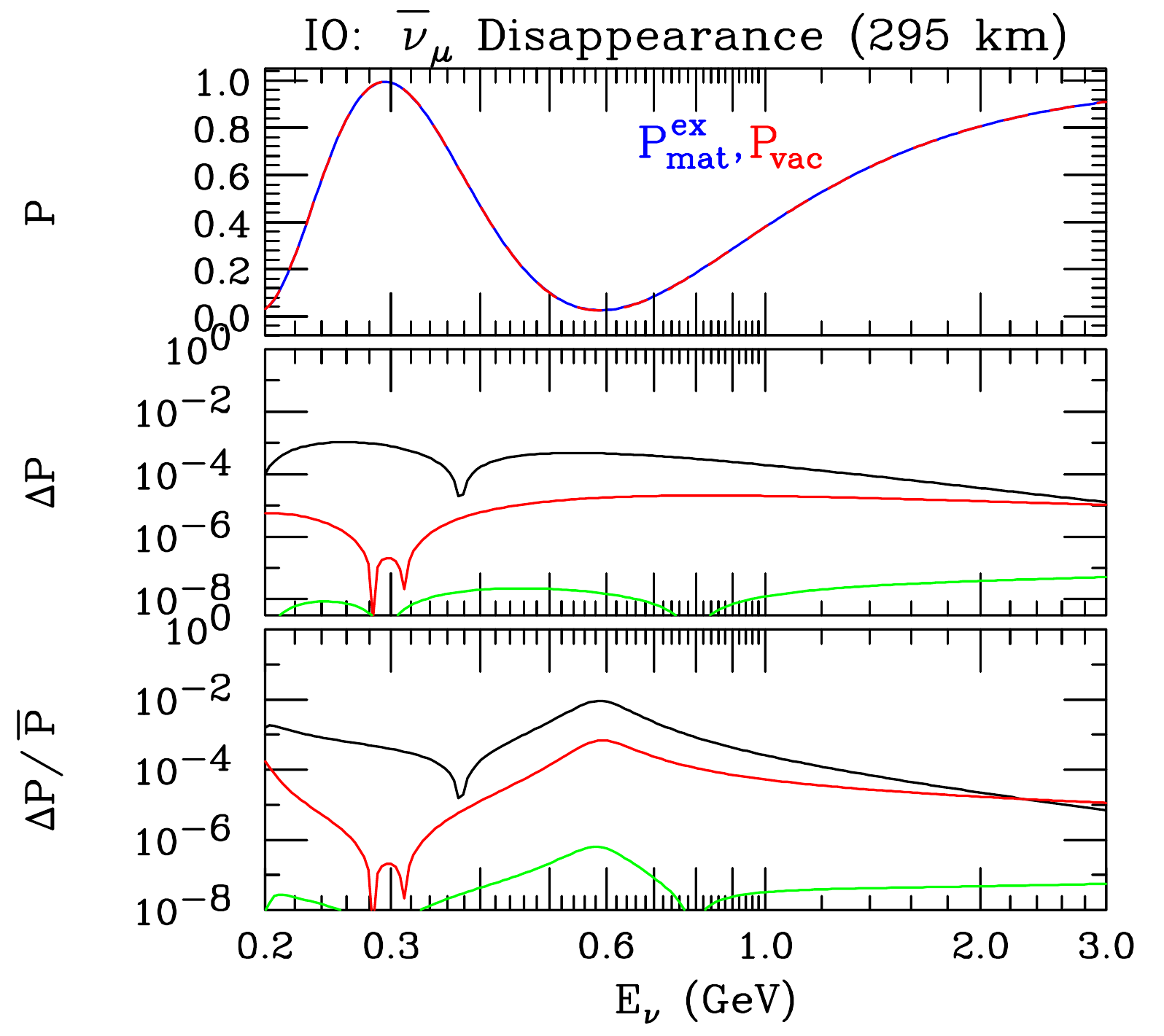}\\[5mm]
     \includegraphics[width=.49\textwidth]{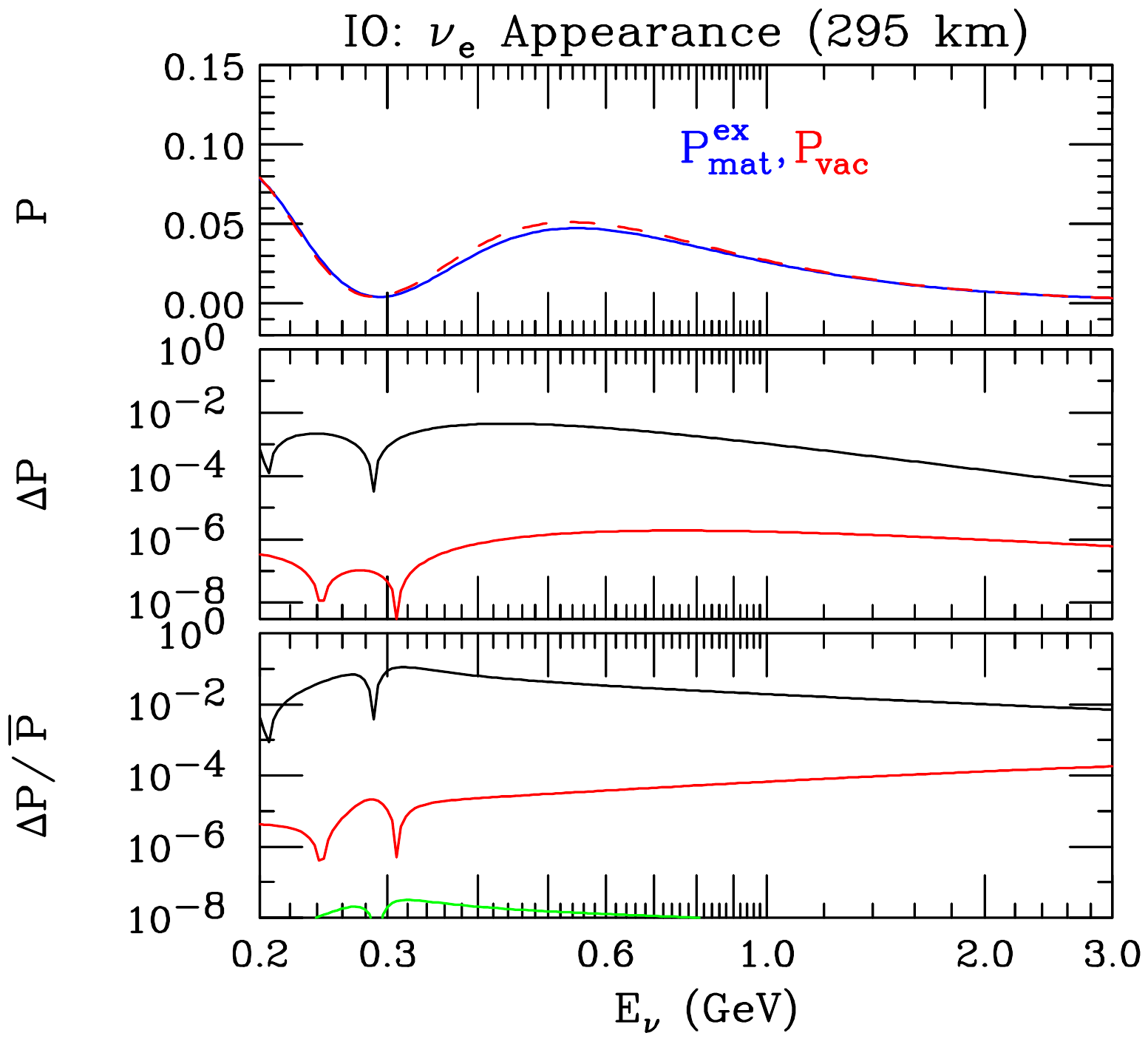}
     \includegraphics[width=.49\textwidth]{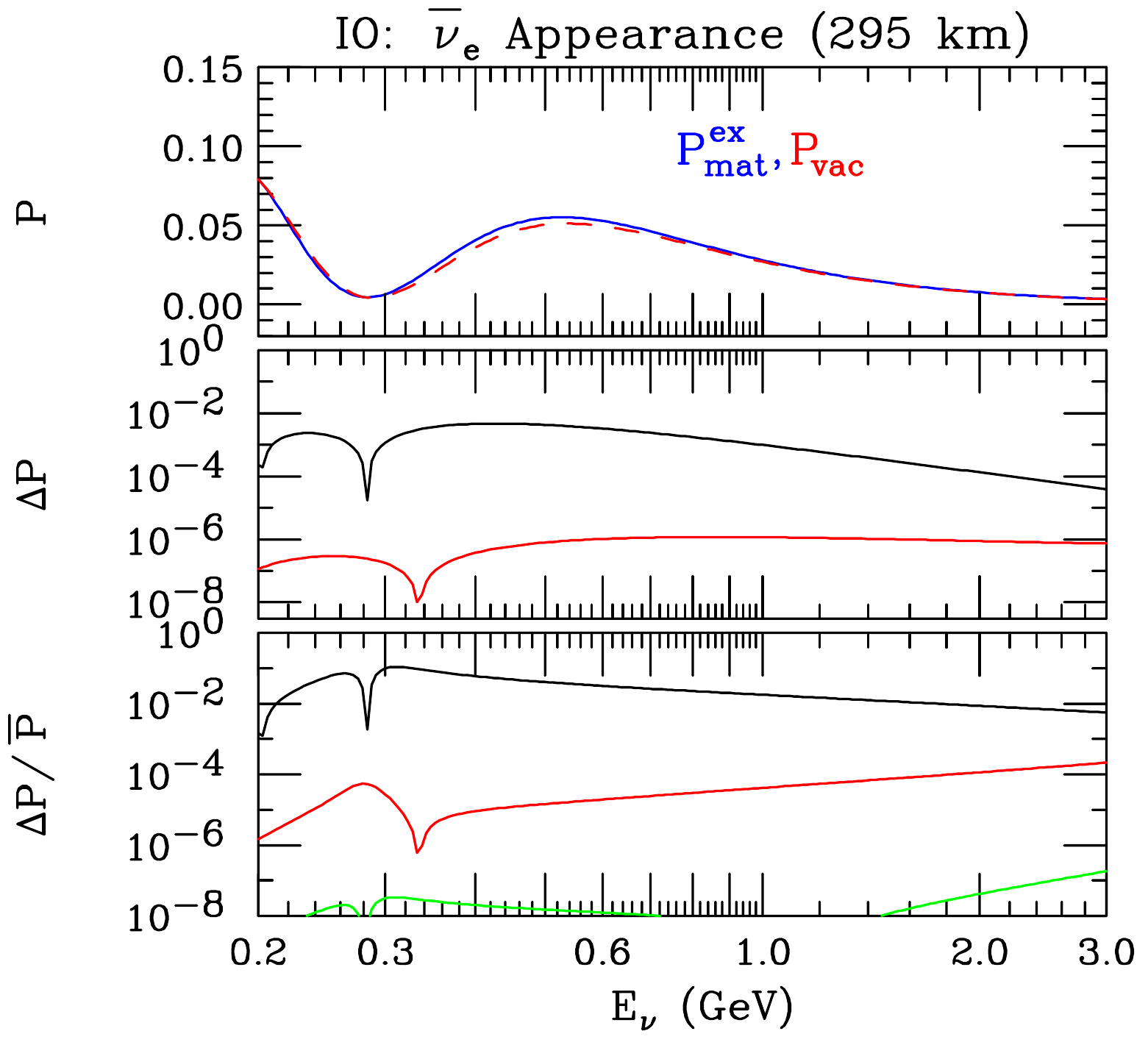}\\[5mm]
             \includegraphics[width=.95\textwidth]{DeltaP.pdf}\\
  \caption{T2K, for inverted ordering (IO): Top Left figure is $\nu_\mu$ disappearance, Top Right figure is $\bar{\nu}_\mu$ disappearance, Bottom Left figure is $\nu_\mu \rightarrow \nu_e$ appearance, and Bottom Right is  $\bar{\nu}_\mu \rightarrow \bar{\nu}_e$ appearance.
  In each figure, the top panel is exact oscillation probability in matter , $P^{ex}_{mat}$,  from \cite{Zaglauer:1988gz}, and in vacuum, $P_{vac}$. The Middle panel is difference between exact oscillation probabilities in matter and vacuum (black), and the difference between exact and 0th (red) and exact and 1st (green) approximations to the matter probabilities  using the DMP scheme, \cite{Denton:2016wmg}. Bottom panel is similar to middle panel but plotting the fractional differences, $\Delta P/\overline{P}$. The density use is 2.3 g.cm$^{-3}$.
  }

     \label{fig:T2K_IO}
          \end{center}

     \end{figure}

\begin{figure}[t]
\begin{center}
     \includegraphics[width=.3\textwidth]{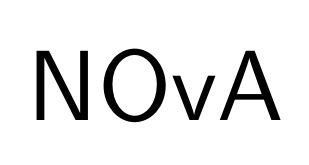}\\
     \includegraphics[width=.49\textwidth]{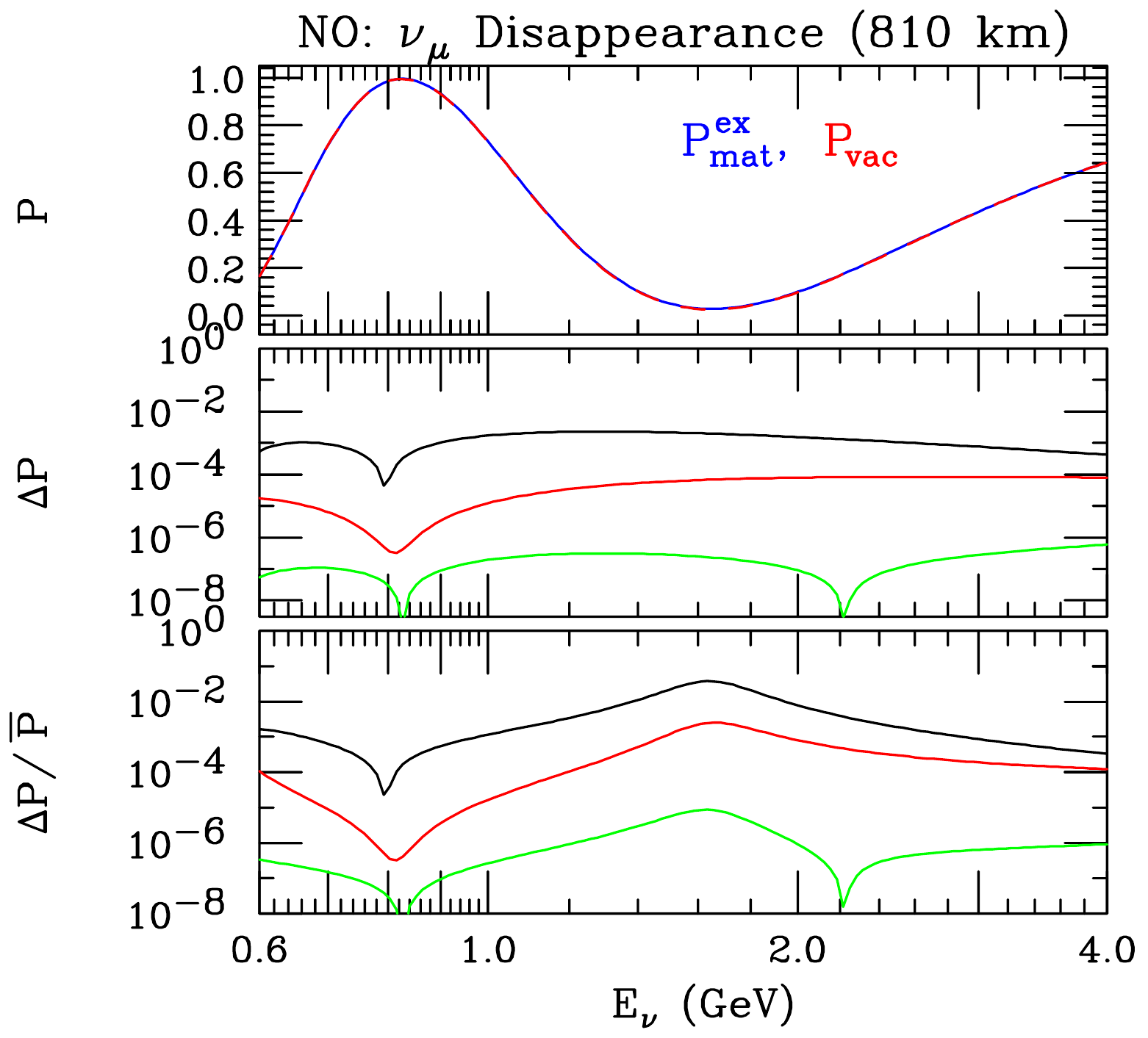}
     \includegraphics[width=.49\textwidth]{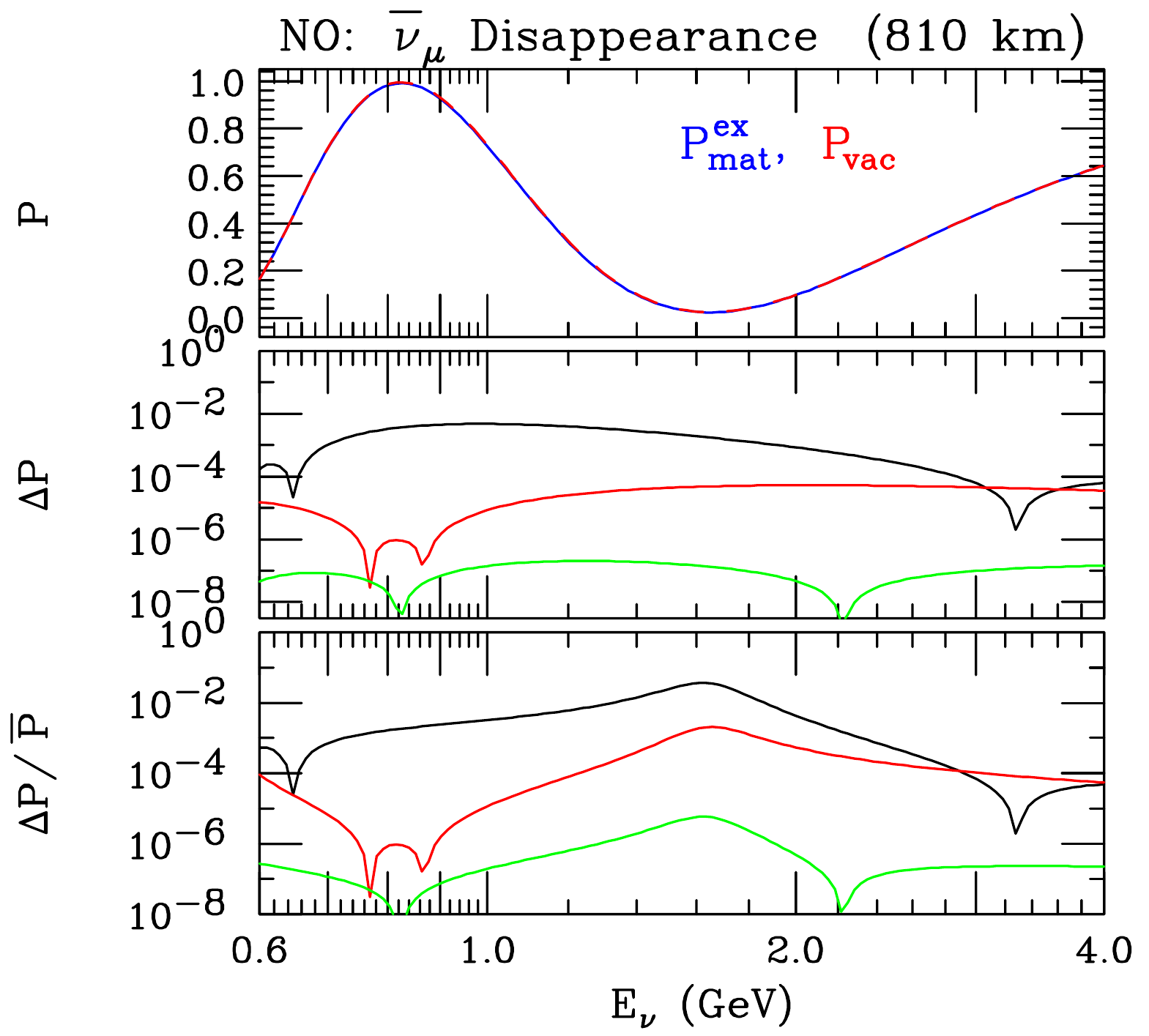}\\[5mm]
     \includegraphics[width=.49\textwidth]{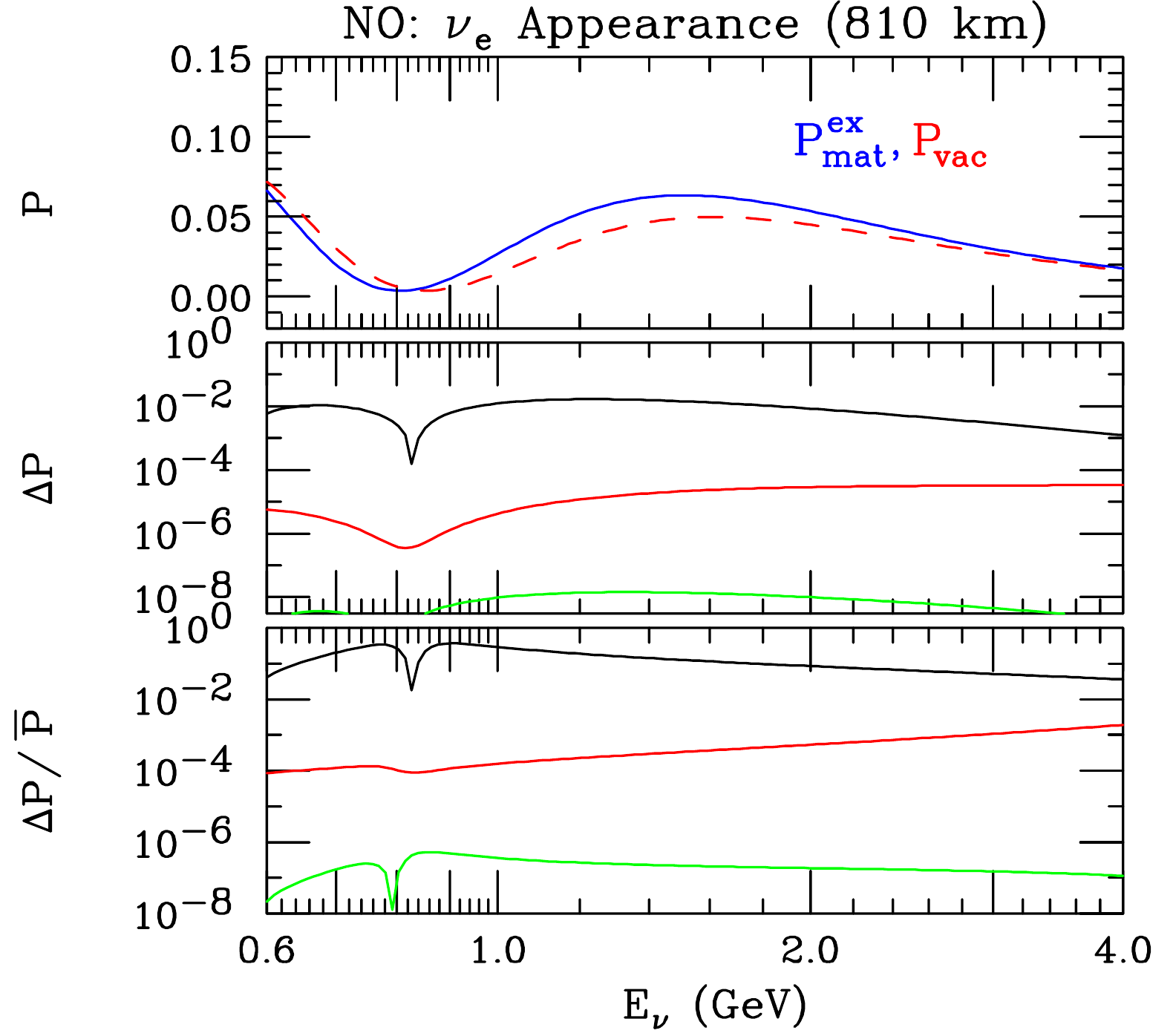}
     \includegraphics[width=.49\textwidth]{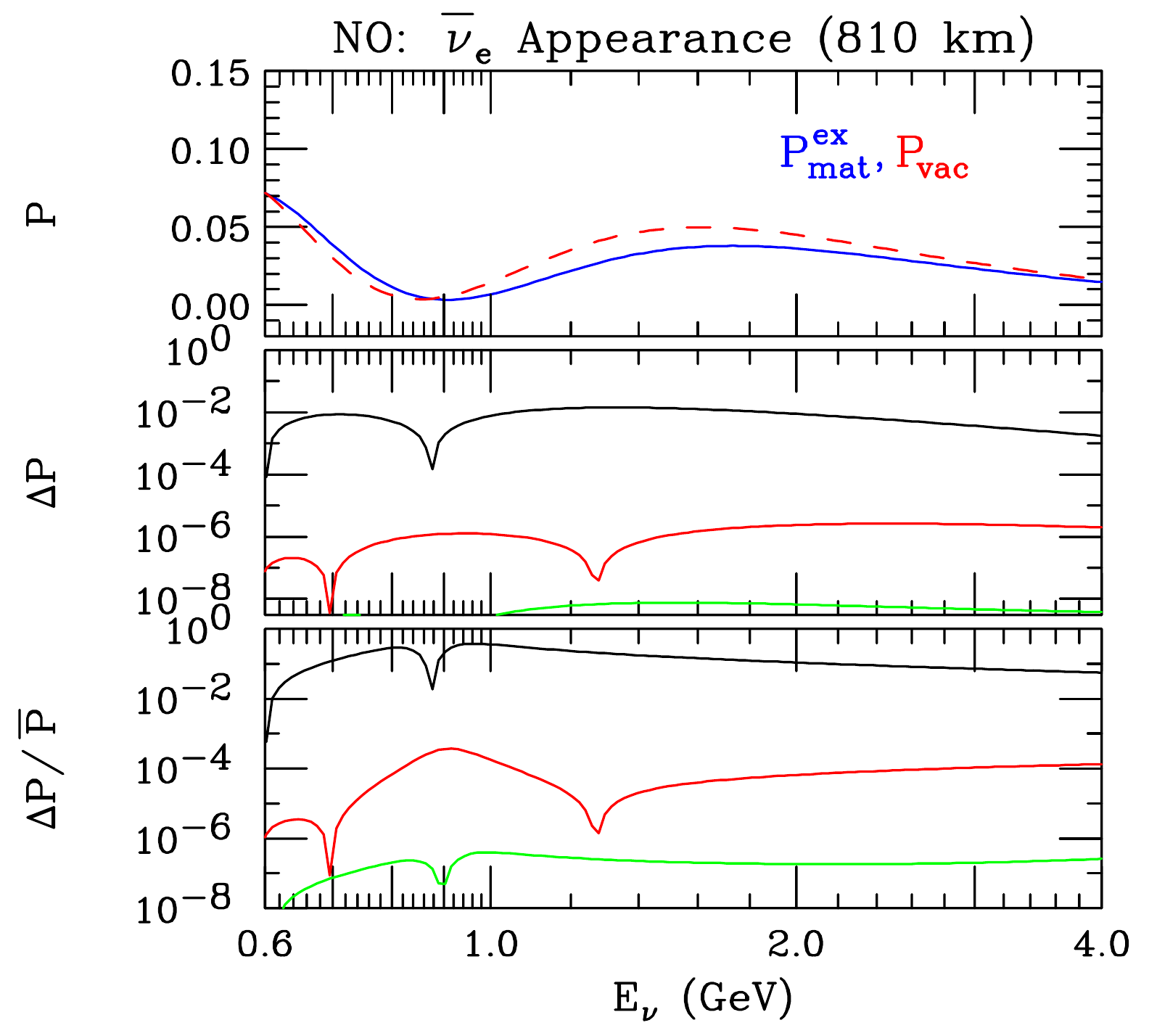}\\[5mm]
             \includegraphics[width=.95\textwidth]{DeltaP.pdf}\\
  \caption{NOvA, for normal ordering (NO): Top Left figure is $\nu_\mu$ disappearance, Top Right figure is $\bar{\nu}_\mu$ disappearance, Bottom Left figure is $\nu_\mu \rightarrow \nu_e$ appearance, and Bottom Right is  $\bar{\nu}_\mu \rightarrow \bar{\nu}_e$ appearance.
  In each figure, the top panel is exact oscillation probability in matter , $P^{ex}_{mat}$,  from \cite{Zaglauer:1988gz}, and in vacuum, $P_{vac}$. The Middle panel is difference between exact oscillation probabilities in matter and vacuum (black), and the difference between exact and 0th (red) and exact and 1st (green) approximations to the matter probabilities  using the DMP scheme, \cite{Denton:2016wmg}.  Bottom panel is similar to middle panel but plotting the fractional differences, $\Delta P/\overline{P}$. The density use is 3.0 g.cm$^{-3}$.
  }
     \label{fig:NOvA_NO}
          \end{center}

     \end{figure}

\begin{figure}[t]
\begin{center}
     \includegraphics[width=.3\textwidth]{NOvA.pdf}\\
     \includegraphics[width=.49\textwidth]{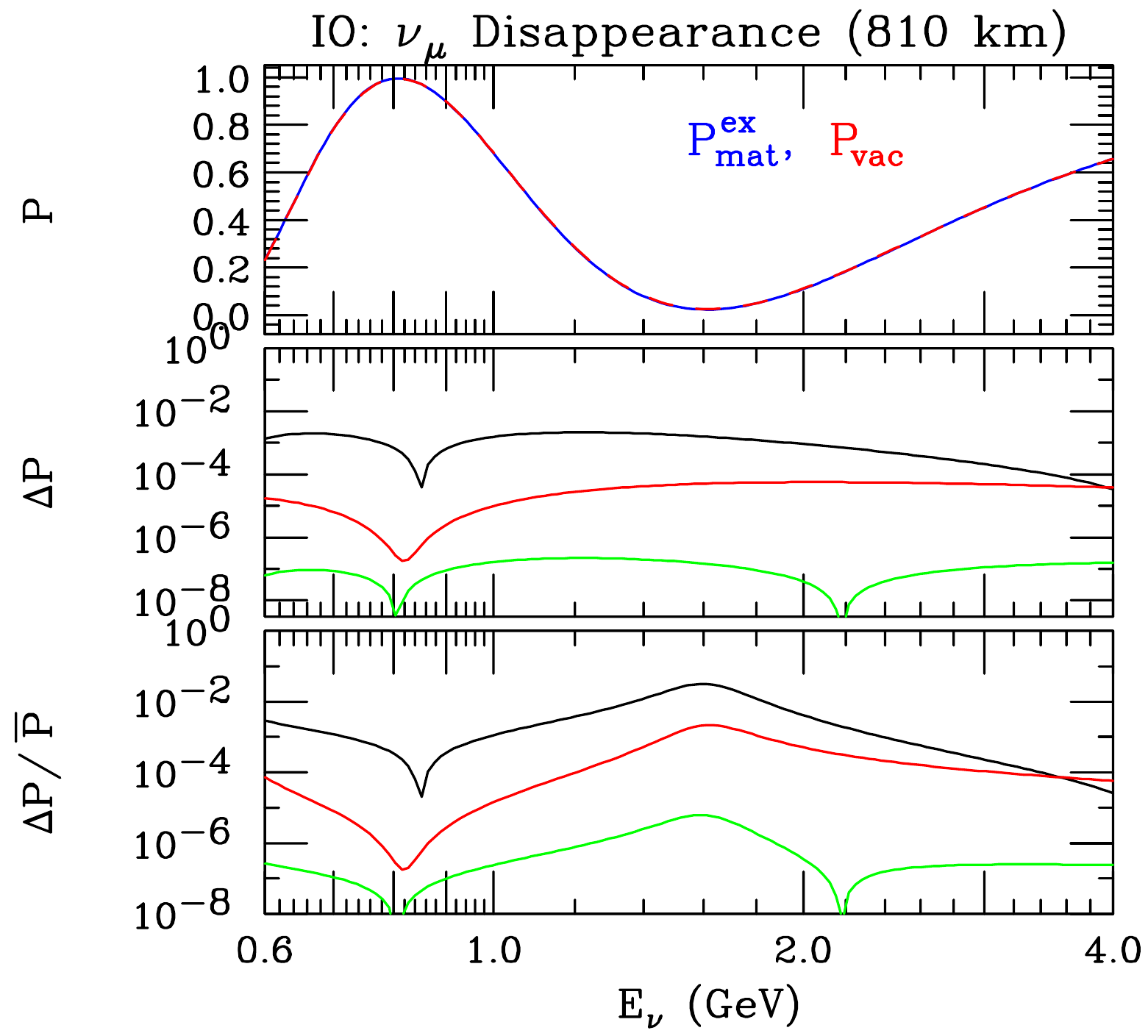}
     \includegraphics[width=.49\textwidth]{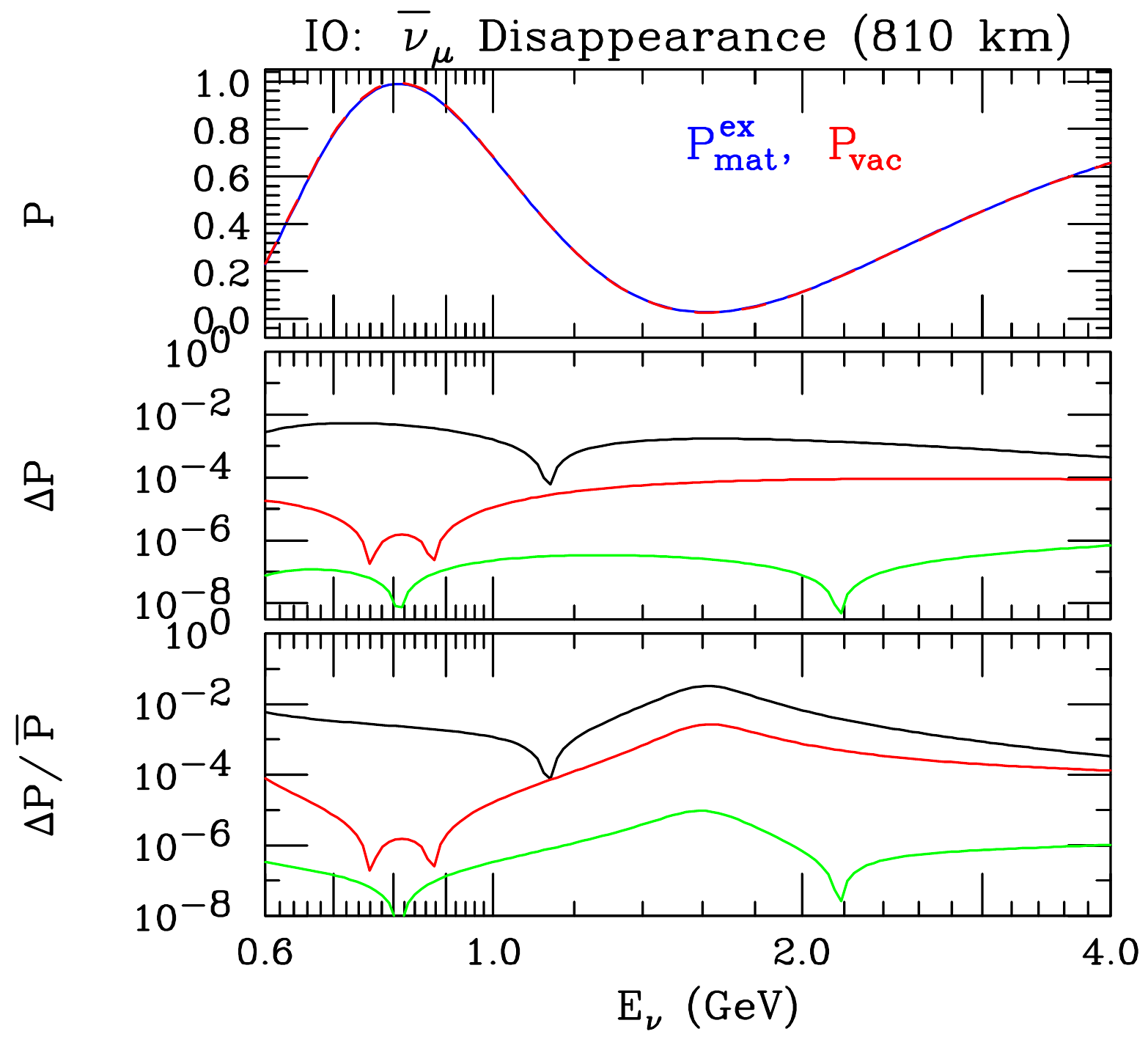}\\[5mm]
     \includegraphics[width=.49\textwidth]{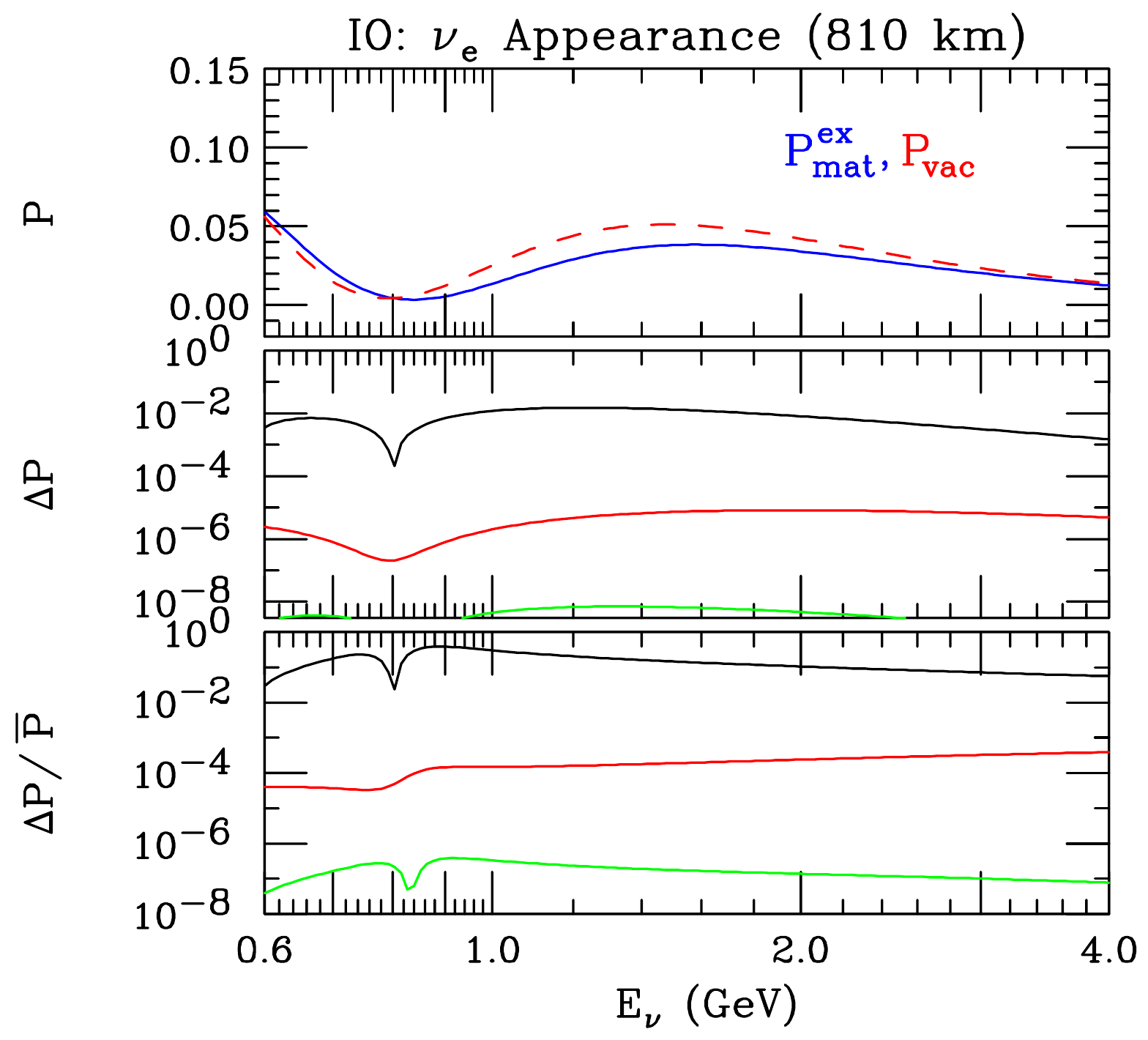}
     \includegraphics[width=.49\textwidth]{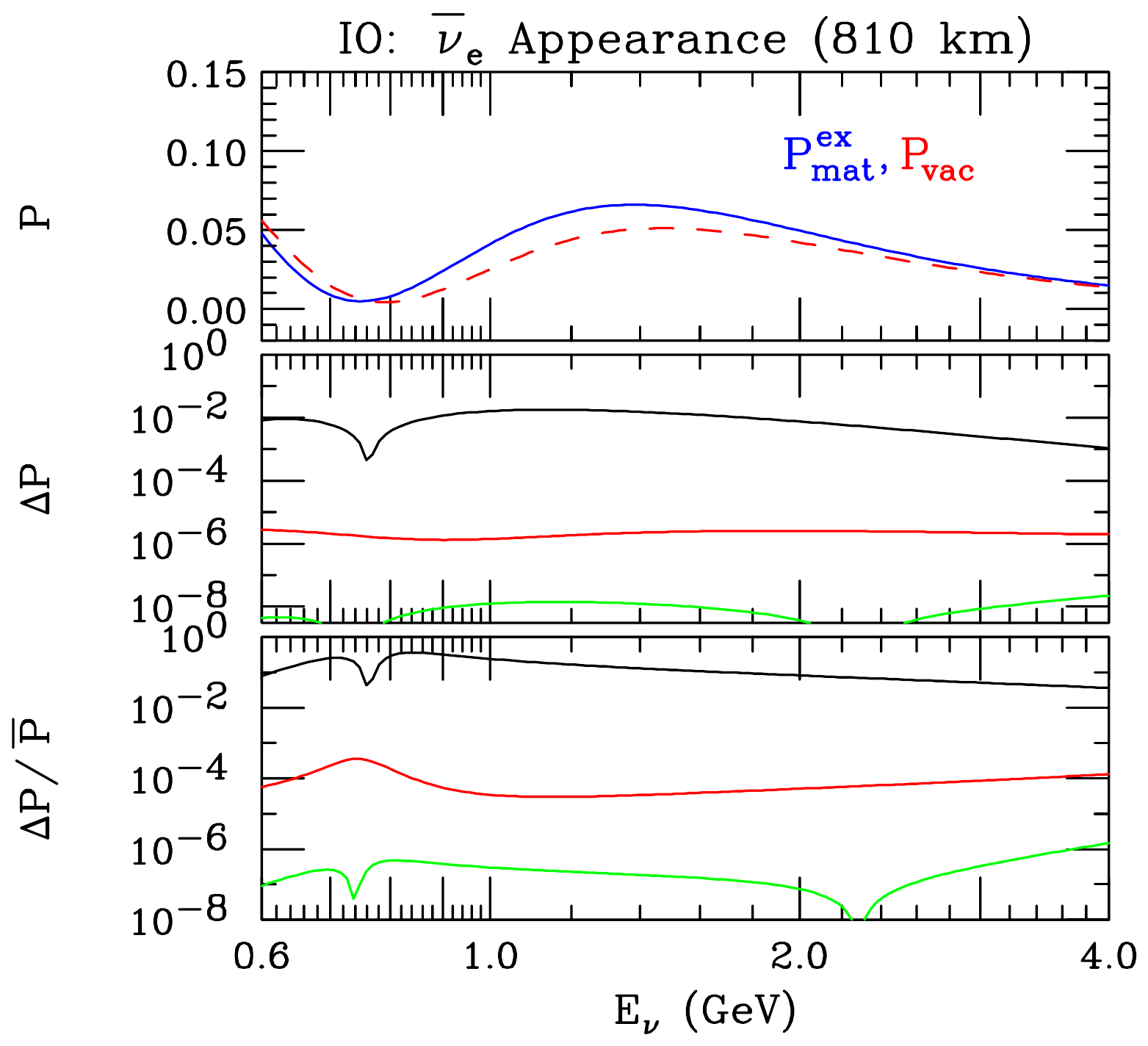}\\[5mm]
             \includegraphics[width=.95\textwidth]{DeltaP.pdf}\\
  \caption{NOvA, for inverted ordering (IO): Top Left figure is $\nu_\mu$ disappearance, Top Right figure is $\bar{\nu}_\mu$ disappearance, Bottom Left figure is $\nu_\mu \rightarrow \nu_e$ appearance, and Bottom Right is  $\bar{\nu}_\mu \rightarrow \bar{\nu}_e$ appearance.
  In each figure, the top panel is exact oscillation probability in matter , $P^{ex}_{mat}$,  from \cite{Zaglauer:1988gz}, and in vacuum, $P_{vac}$. The Middle panel is difference between exact oscillation probabilities in matter and vacuum (black), and the difference between exact and 0th (red) and exact and 1st (green) approximations to the matter probabilities  using the DMP scheme, \cite{Denton:2016wmg}. Bottom panel is similar to middle panel but plotting the fractional differences, $\Delta P/\overline{P}$. The density use is 3.0 g.cm$^{-3}$.
  }
     \label{fig:NOvA_IO}
          \end{center}

     \end{figure}

\begin{figure}[t]
\begin{center}
     \includegraphics[width=.3\textwidth]{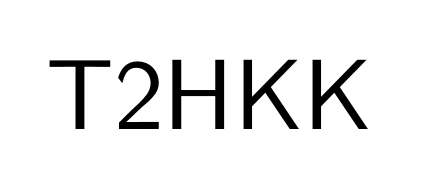}\\
     \includegraphics[width=.49\textwidth]{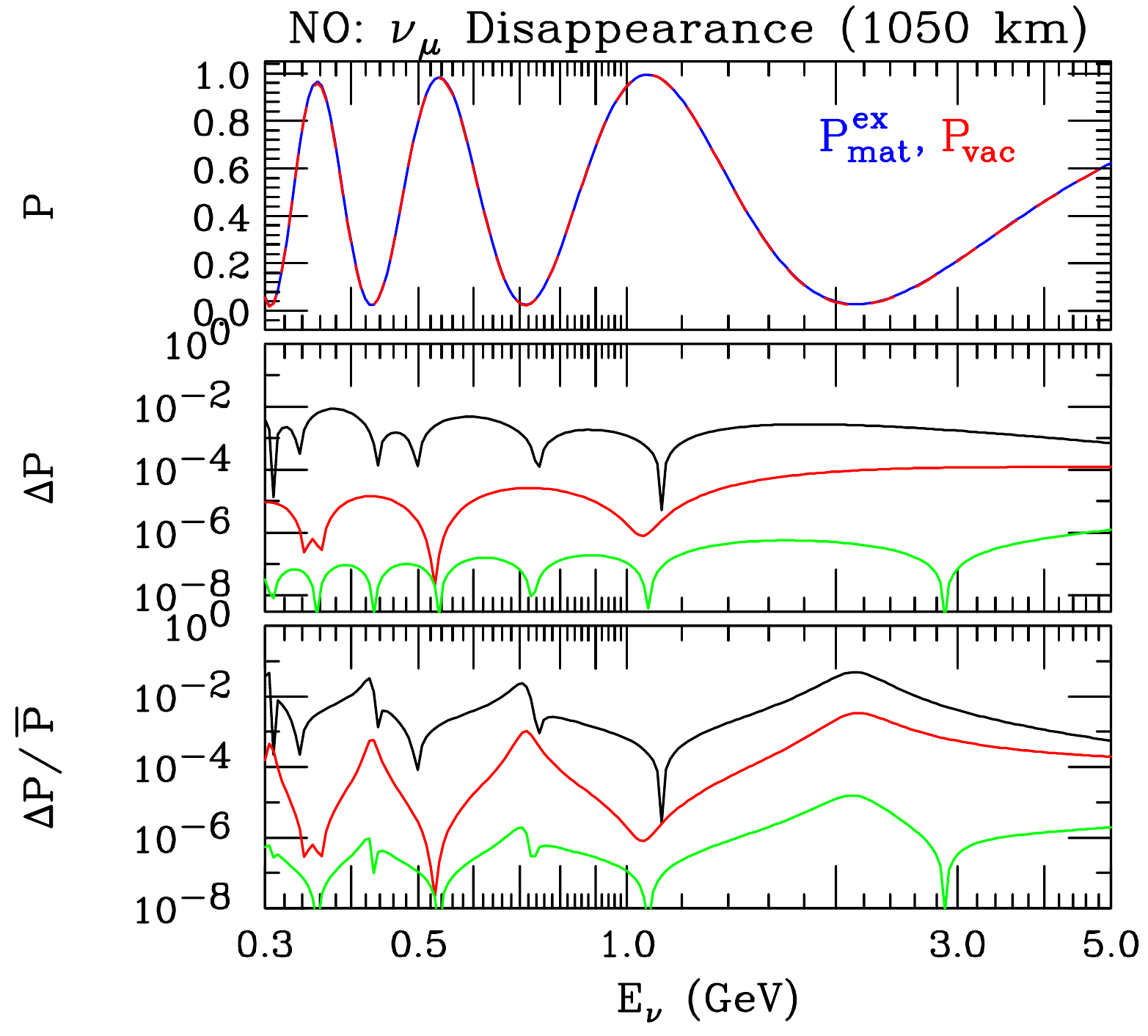}
     \includegraphics[width=.49\textwidth]{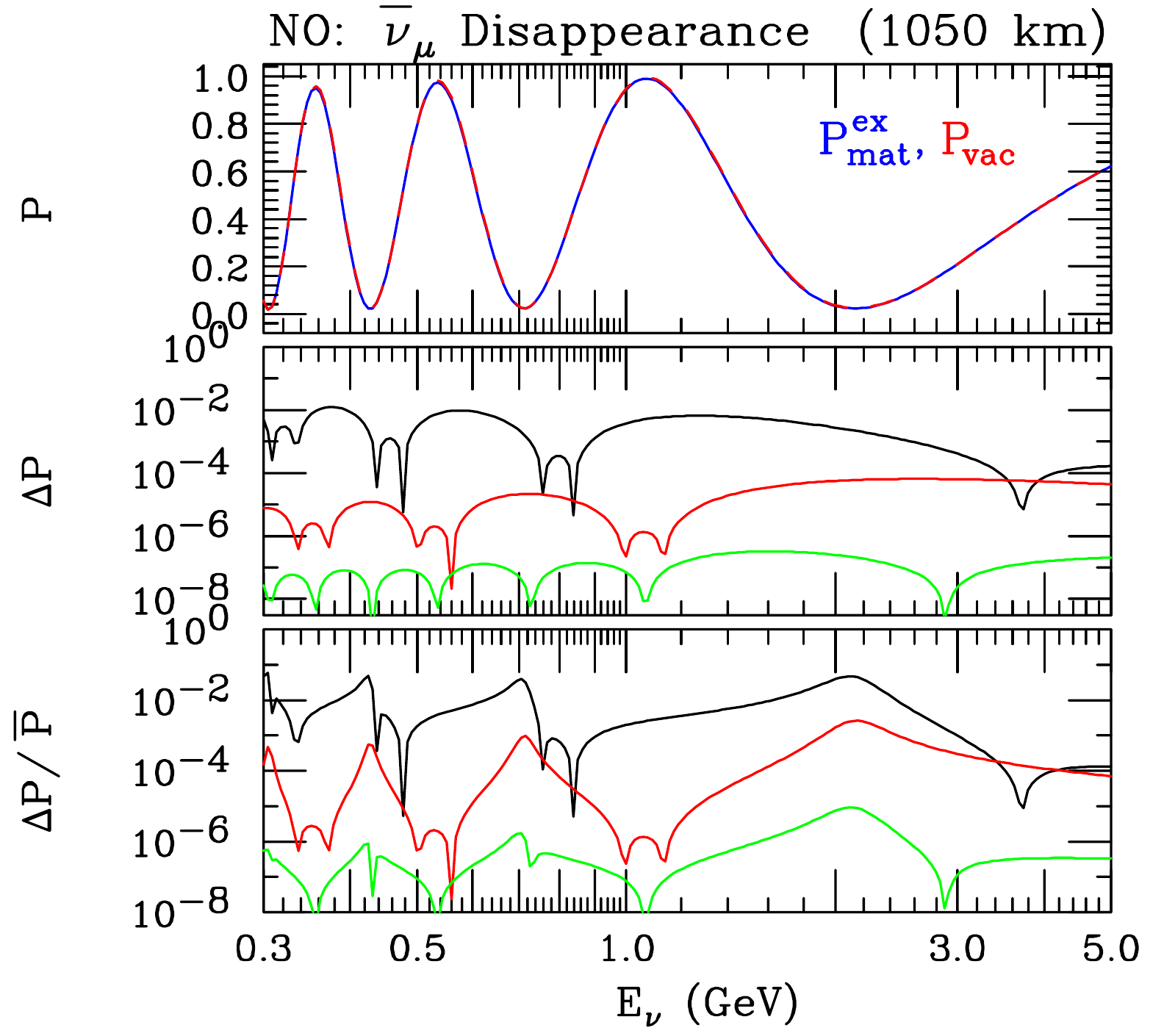}\\[5mm]
     \includegraphics[width=.49\textwidth]{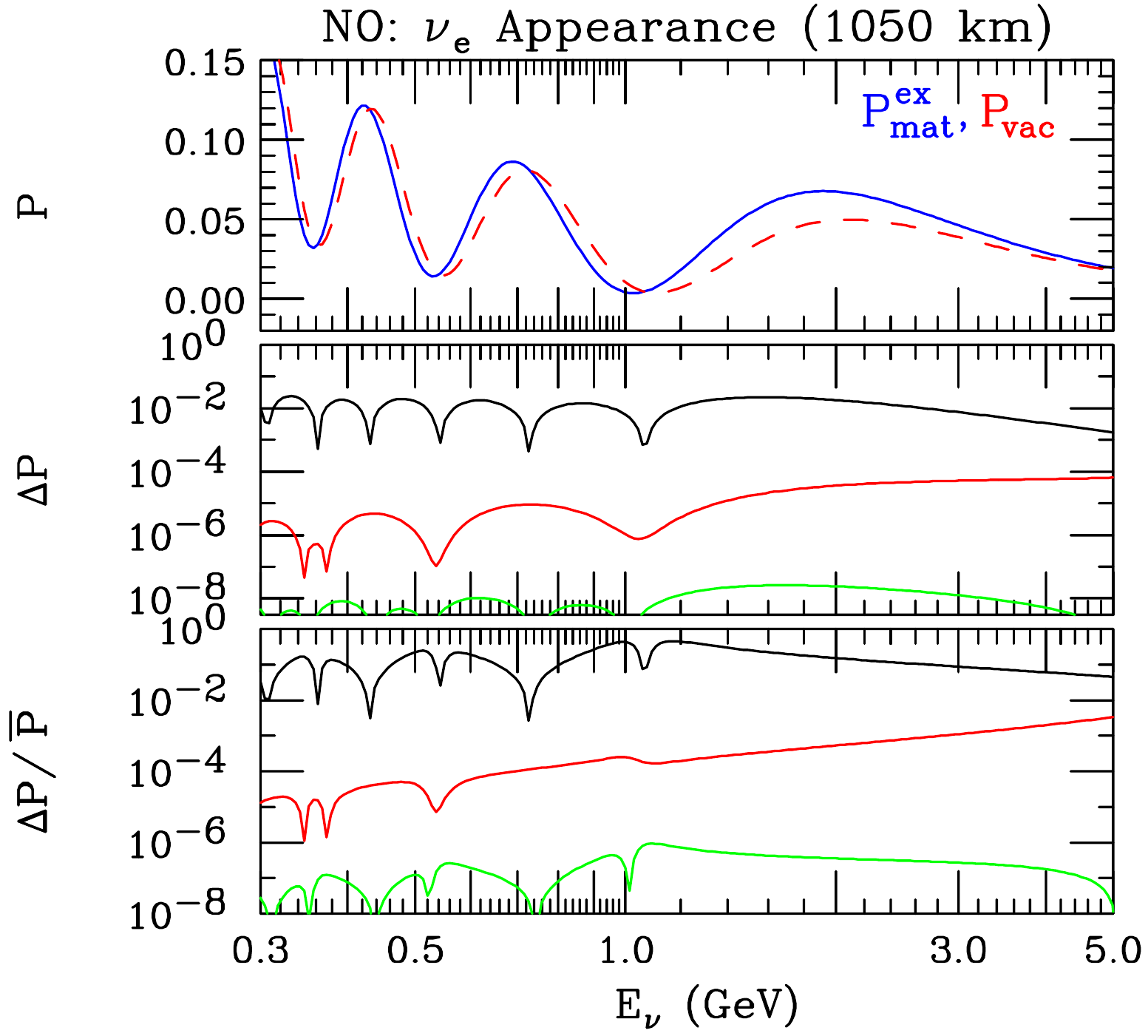}
     \includegraphics[width=.49\textwidth]{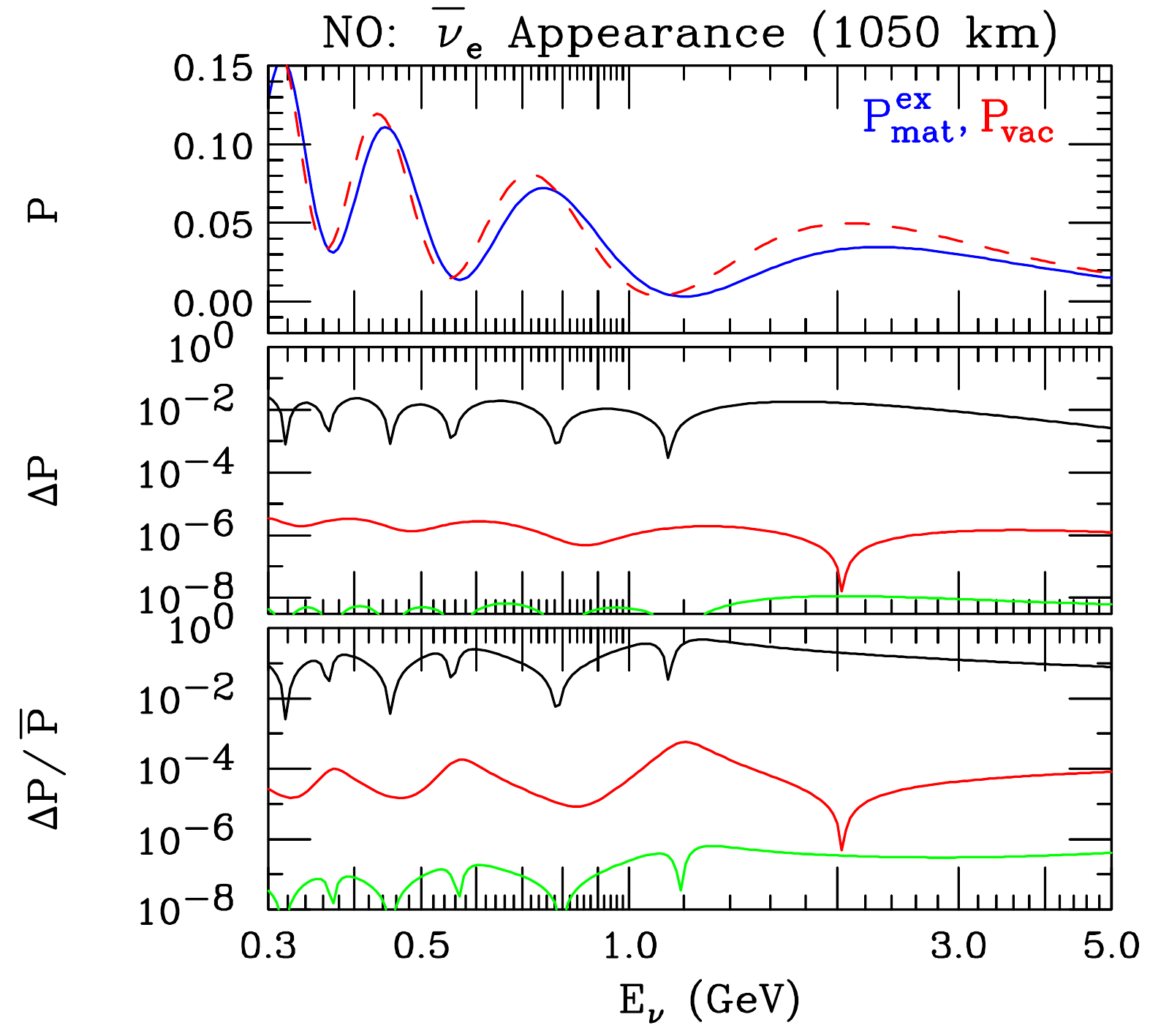}\\[5mm]
             \includegraphics[width=.95\textwidth]{DeltaP.pdf}\\
  \caption{T2HKK, for normal ordering (NO): Top Left figure is $\nu_\mu$ disappearance, Top Right figure is $\bar{\nu}_\mu$ disappearance, Bottom Left figure is $\nu_\mu \rightarrow \nu_e$ appearance, and Bottom Right is  $\bar{\nu}_\mu \rightarrow \bar{\nu}_e$ appearance.
  In each figure, the top panel is exact oscillation probability in matter , $P^{ex}_{mat}$,  from \cite{Zaglauer:1988gz}, and in vacuum, $P_{vac}$. The Middle panel is difference between exact oscillation probabilities in matter and vacuum (black), and the difference between exact and 0th (red) and exact and 1st (green) approximations to the matter probabilities  using the DMP scheme, \cite{Denton:2016wmg}.  Bottom panel is similar to middle panel but plotting the fractional differences, $\Delta P/\overline{P}$. The density use is 3.0 g.cm$^{-3}$.
  }
     \label{fig:T2HKK_NO}
          \end{center}

     \end{figure}

\begin{figure}[b]
\begin{center}
   \includegraphics[width=.3\textwidth]{T2HKK.pdf}\\
     \includegraphics[width=.49\textwidth]{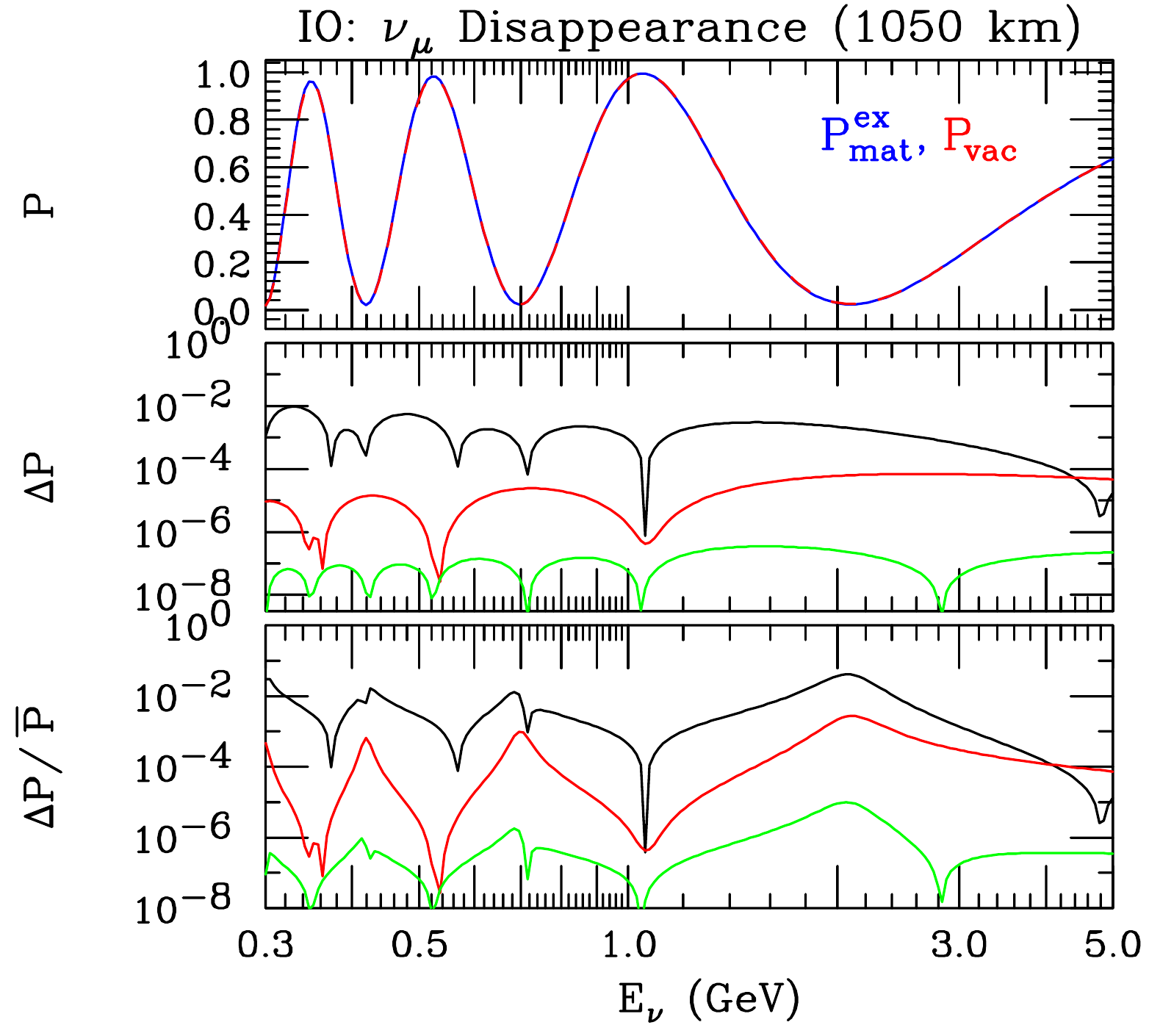}
     \includegraphics[width=.49\textwidth]{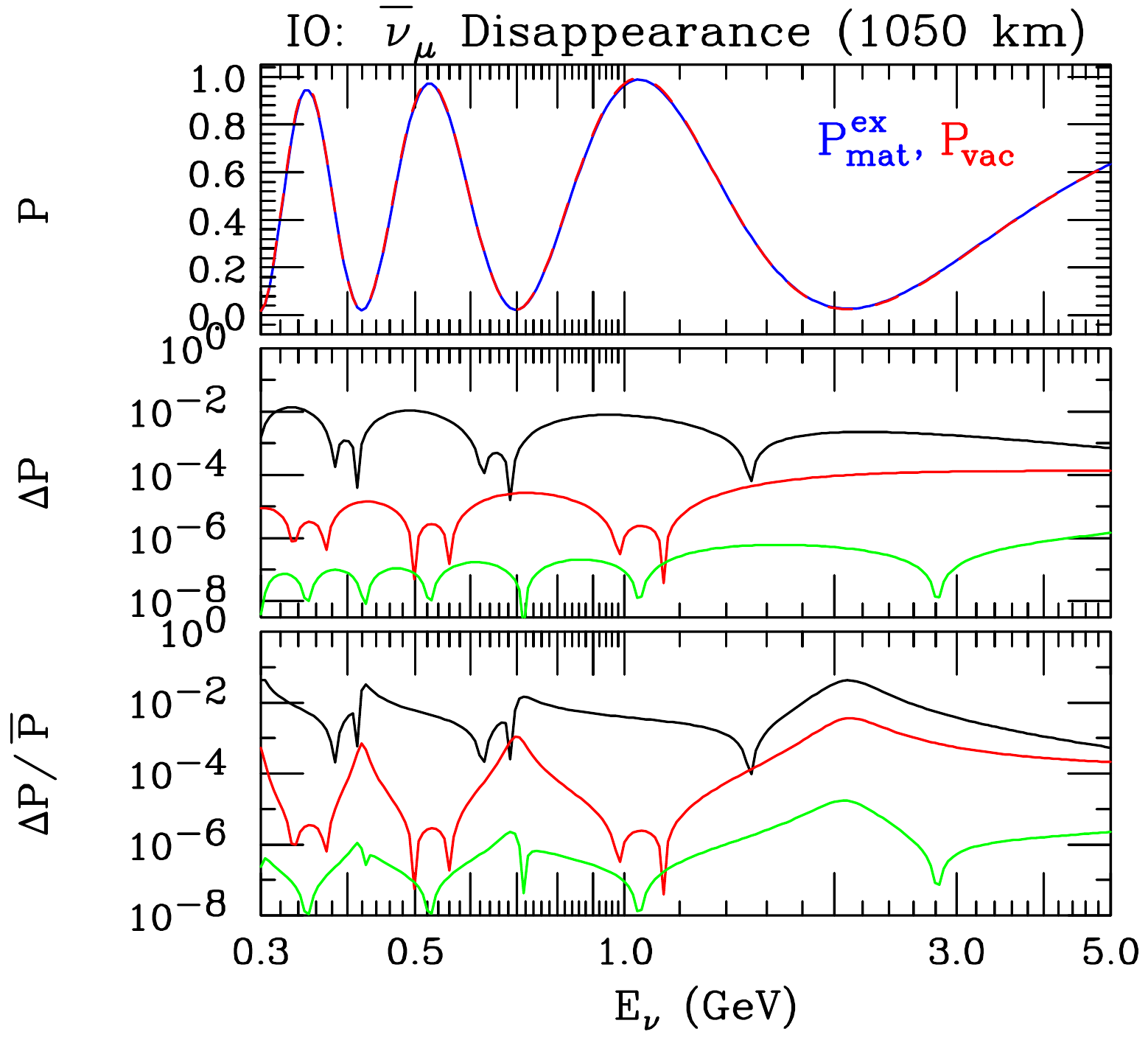}\\[5mm]
     \includegraphics[width=.49\textwidth]{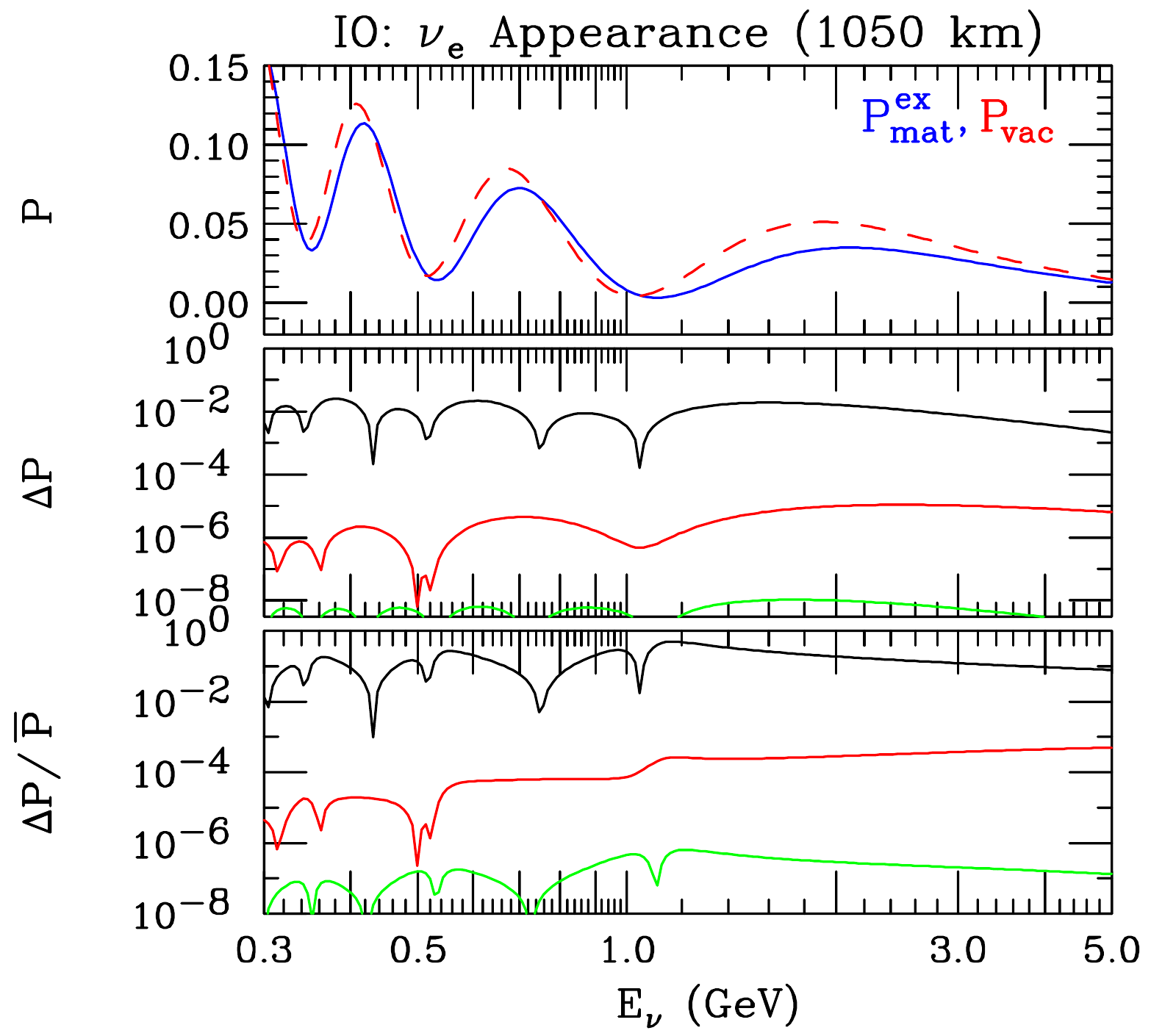}
     \includegraphics[width=.49\textwidth]{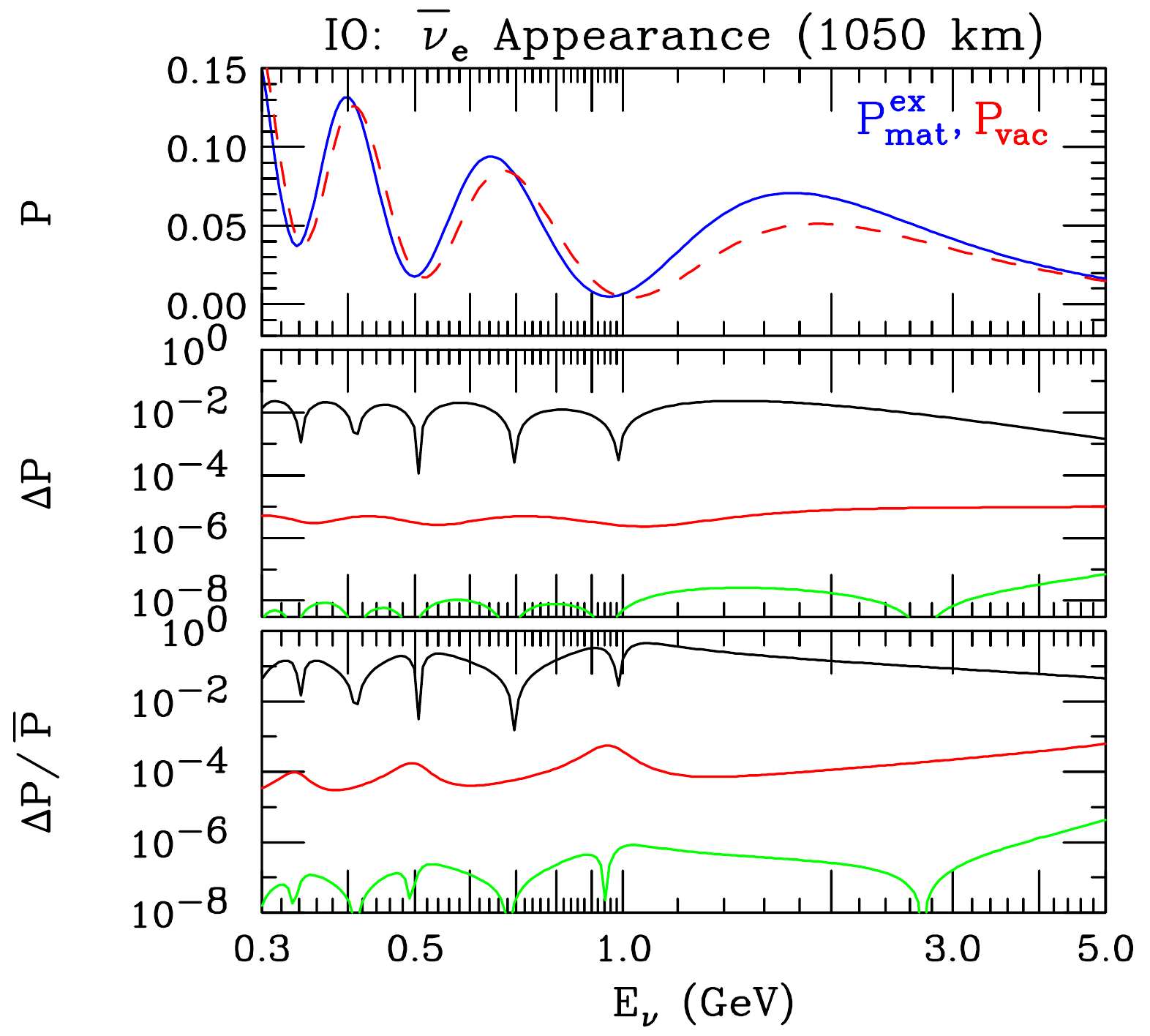}\\[2mm]
             \includegraphics[width=.95\textwidth]{DeltaP.pdf}\\
  \caption{T2HKK, for inverted ordering (IO): Top Left figure is $\nu_\mu$ disappearance, Top Right figure is $\bar{\nu}_\mu$ disappearance, Bottom Left figure is $\nu_\mu \rightarrow \nu_e$ appearance, and Bottom Right is  $\bar{\nu}_\mu \rightarrow \bar{\nu}_e$ appearance.
  In each figure, the top panel is exact oscillation probability in matter , $P^{ex}_{mat}$, from \cite{Zaglauer:1988gz}, and in vacuum, $P_{vac}$. The Middle panel is difference between exact oscillation probabilities in matter and vacuum (black), and the difference between exact and 0th (red) and exact and 1st (green) approximations to the matter probabilities  using the DMP scheme, \cite{Denton:2016wmg}.  Bottom panel is similar to middle panel but plotting the fractional differences, $\Delta P/\overline{P}$. The density use is 3.0 g.cm$^{-3}$.
}
      \label{fig:T2HKK_IO}
          \end{center}

     \end{figure}

\begin{figure}[t]
\begin{center}
     \includegraphics[width=.3\textwidth]{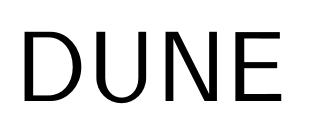}\\
     \includegraphics[width=.49\textwidth]{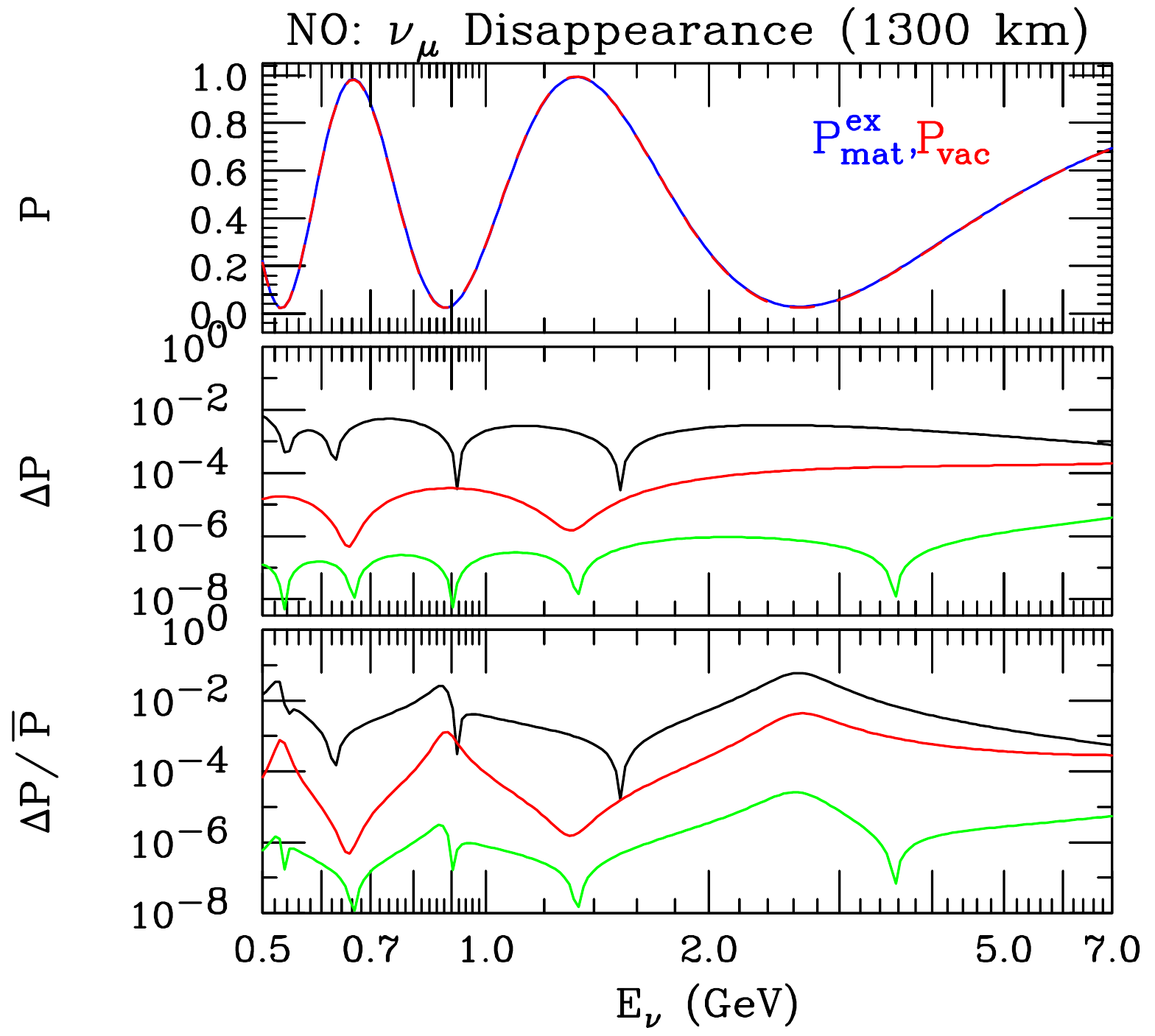}
     \includegraphics[width=.49\textwidth]{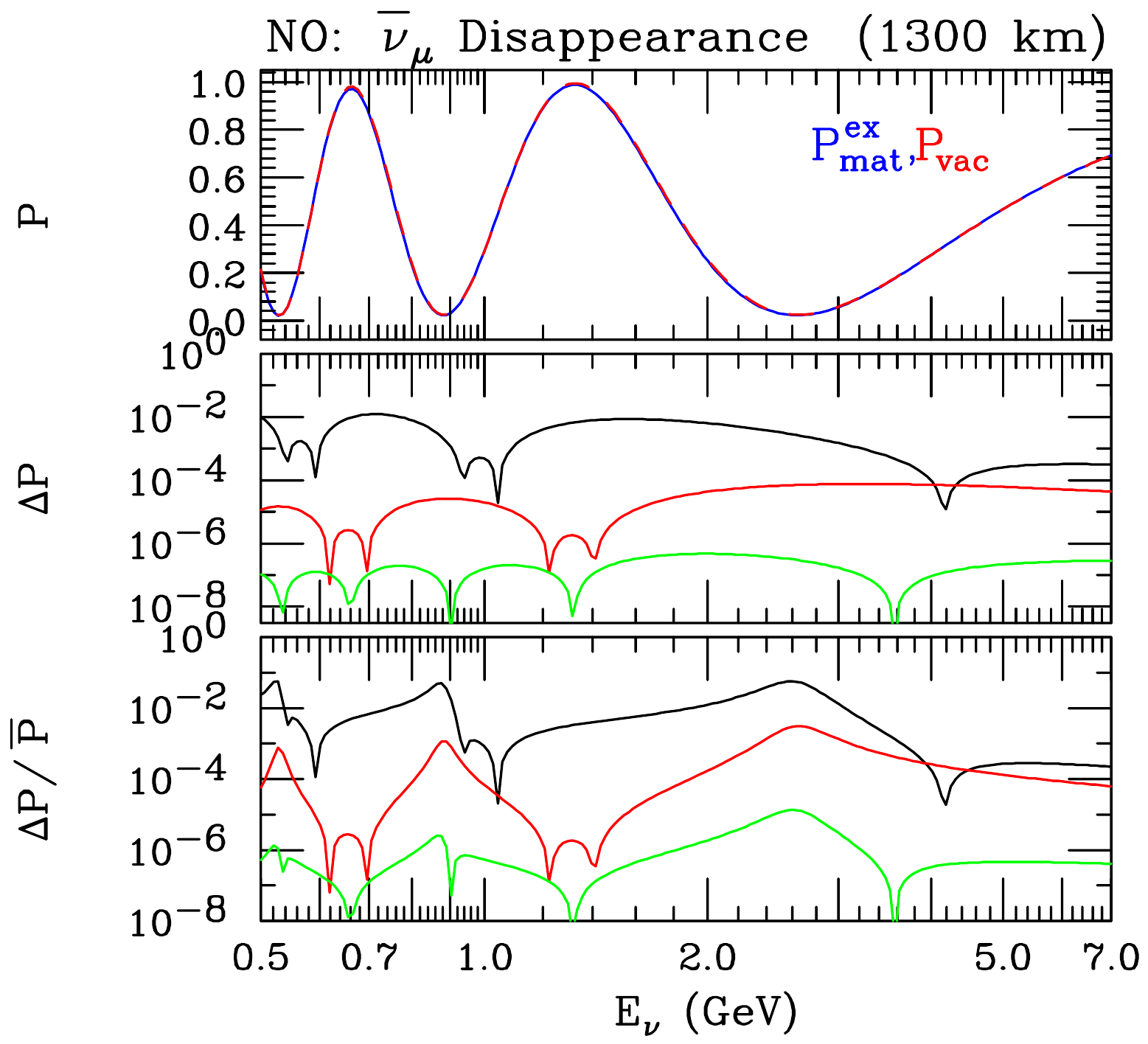}\\[5mm]
     \includegraphics[width=.49\textwidth]{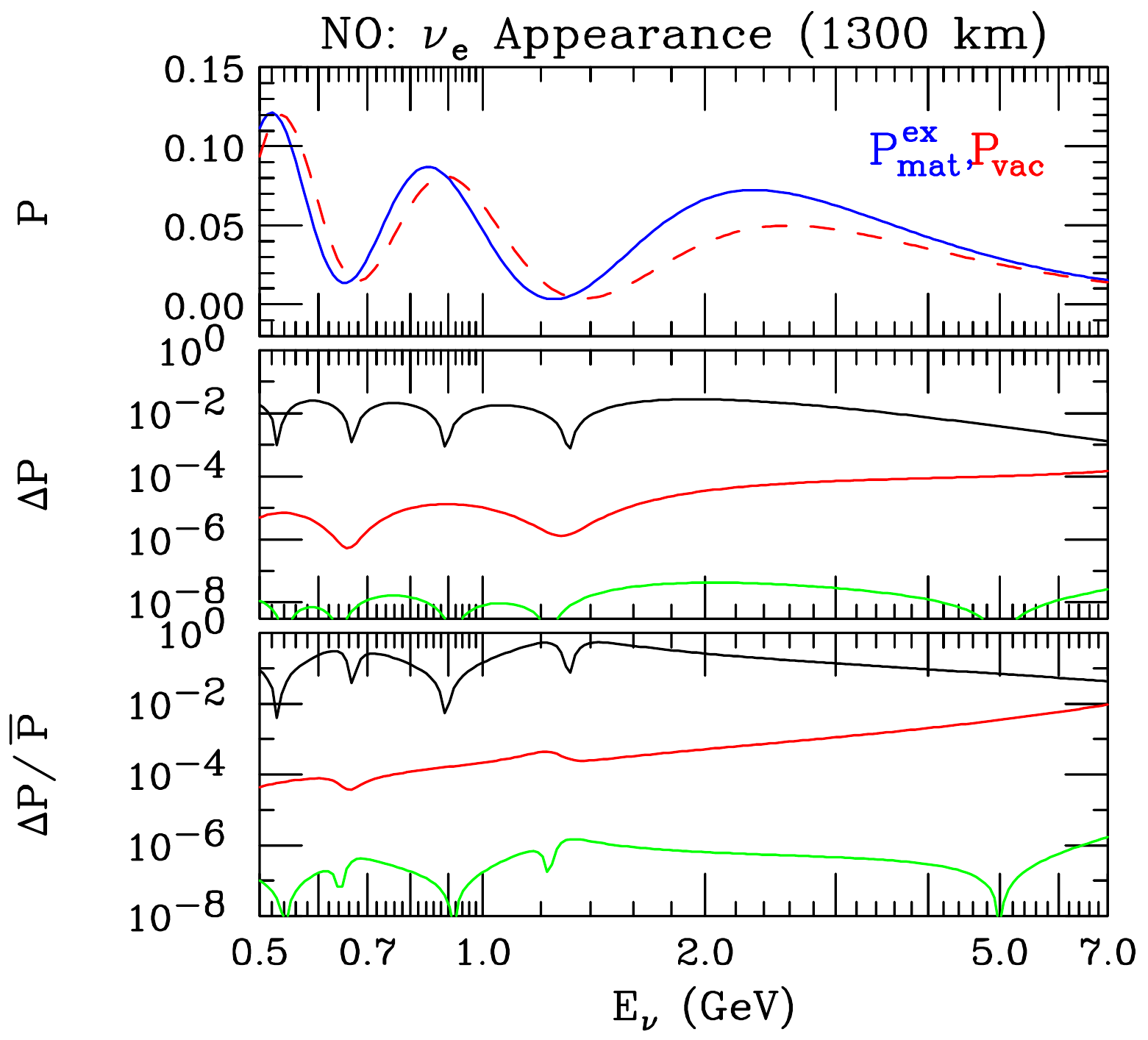}
     \includegraphics[width=.49\textwidth]{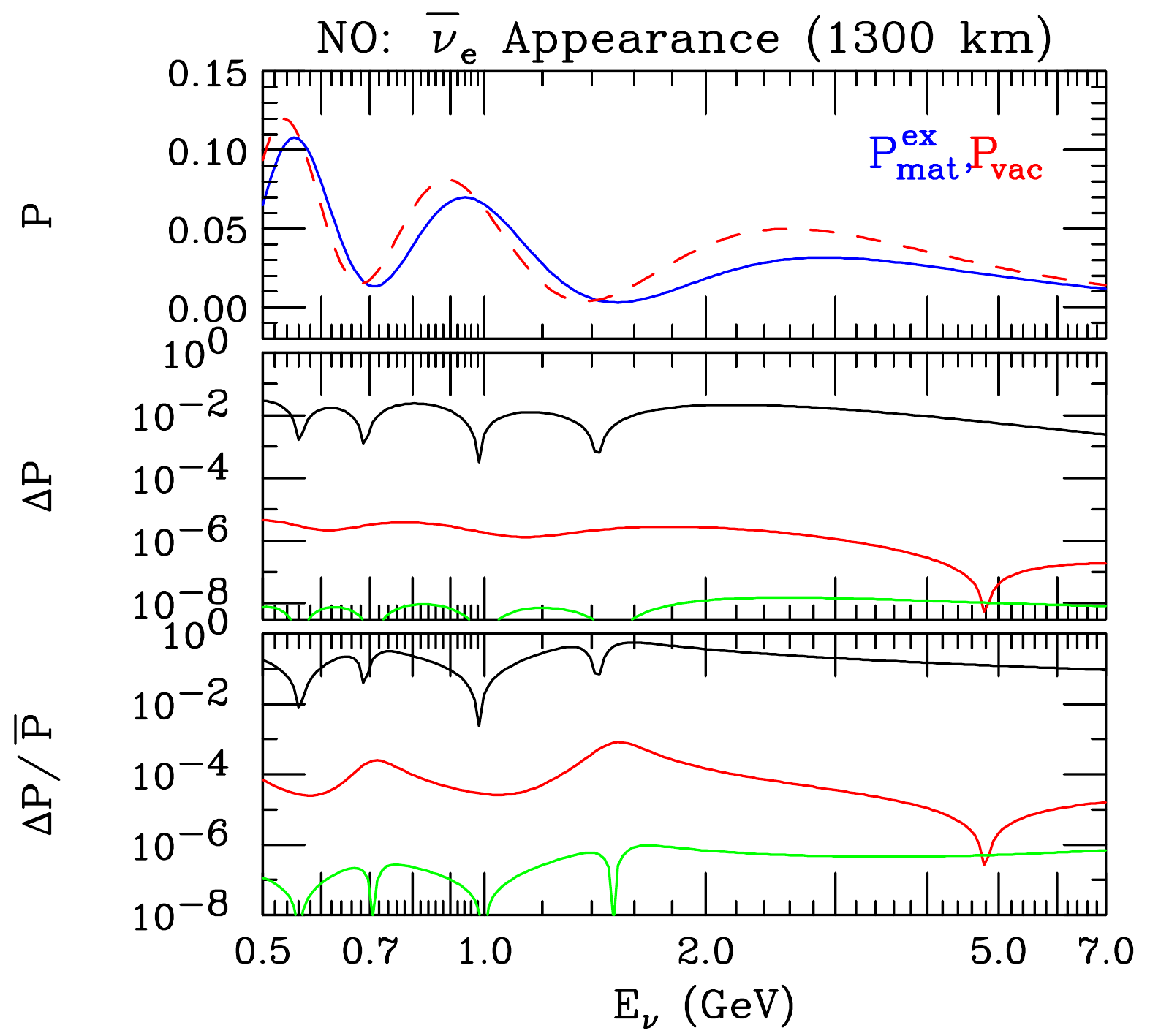}\\[5mm]
             \includegraphics[width=.95\textwidth]{DeltaP.pdf}\\
  \caption{DUNE, for normal ordering (NO): Top Left figure is $\nu_\mu$  disappearance,   Top Right figure is $\bar{\nu}_\mu$ disappearance, Bottom Left figure is $\nu_\mu \rightarrow \nu_e$ appearance, and Bottom Right is  $\bar{\nu}_\mu \rightarrow \bar{\nu}_e$ appearance.
  In each figure, the top panel is exact oscillation probability in matter , $P^{ex}_{mat}$,  from \cite{Zaglauer:1988gz}, and in vacuum, $P_{vac}$. The Middle panel is difference between exact oscillation probabilities in matter and vacuum (black), and the difference,  
  $\Delta P$, between exact and 0th (red) and exact and 1st (green) approximations to the matter probabilities  using the DMP scheme, \cite{Denton:2016wmg}.  Bottom panel is similar to middle panel but plotting the fractional differences, $\Delta P/\overline{P}$. The density use is 3.0 g.cm$^{-3}$.
}
     \label{fig:DUNE_NO}
          \end{center}

     \end{figure}

\begin{figure}[t]
\begin{center}
     \includegraphics[width=.3\textwidth]{DUNE.pdf}\\
     \includegraphics[width=.49\textwidth]{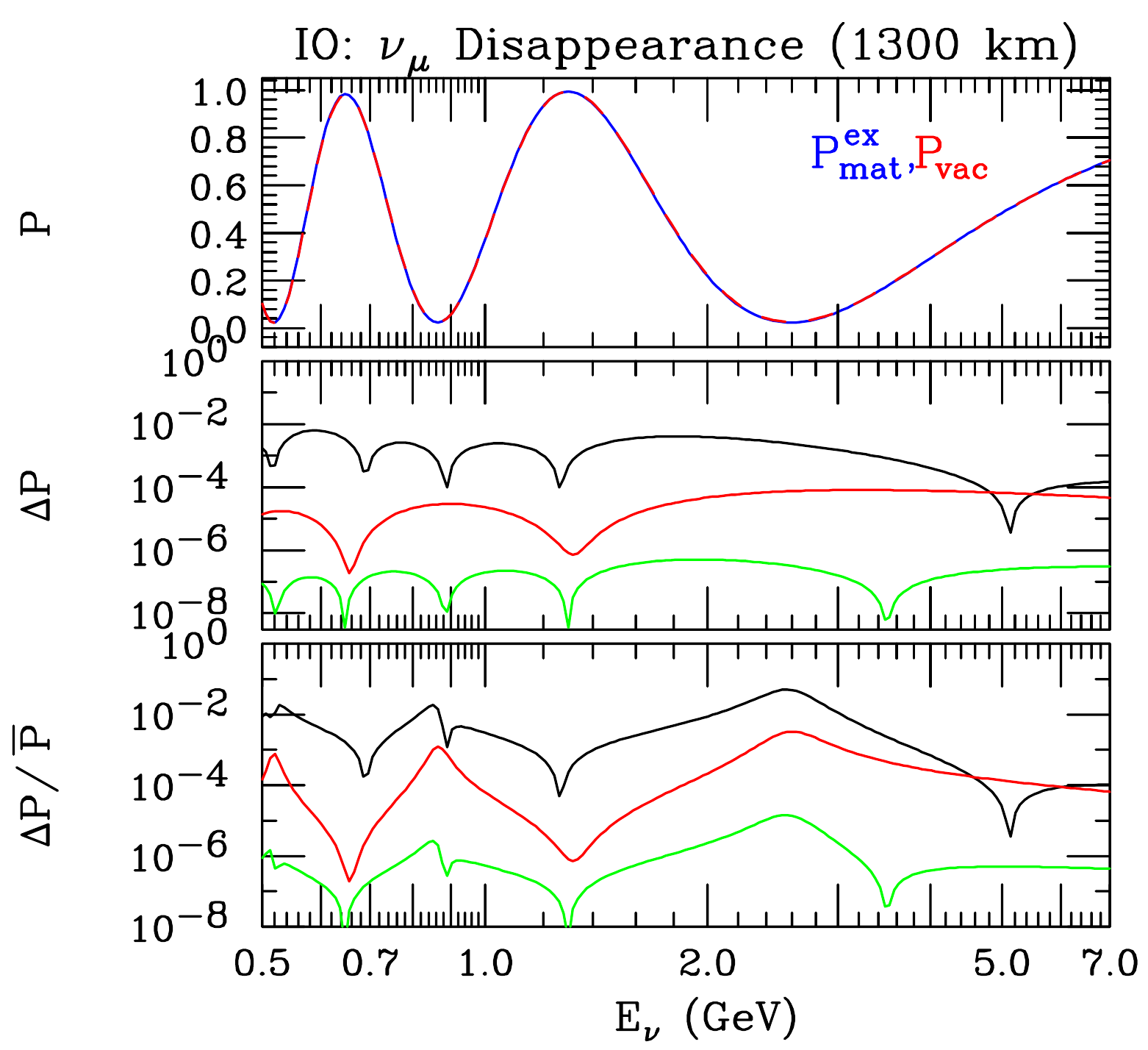}
     \includegraphics[width=.49\textwidth]{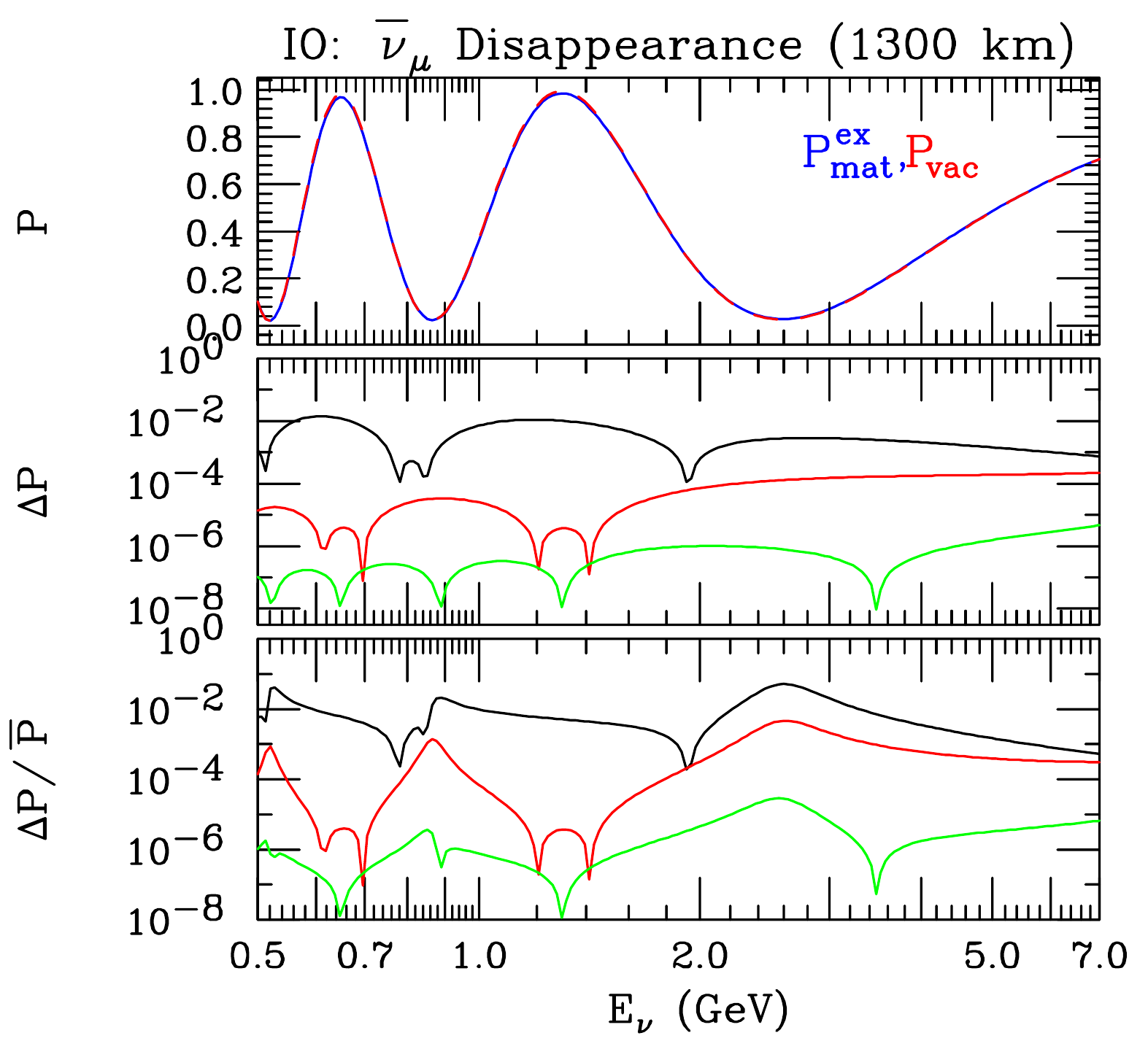}\\[5mm]
     \includegraphics[width=.49\textwidth]{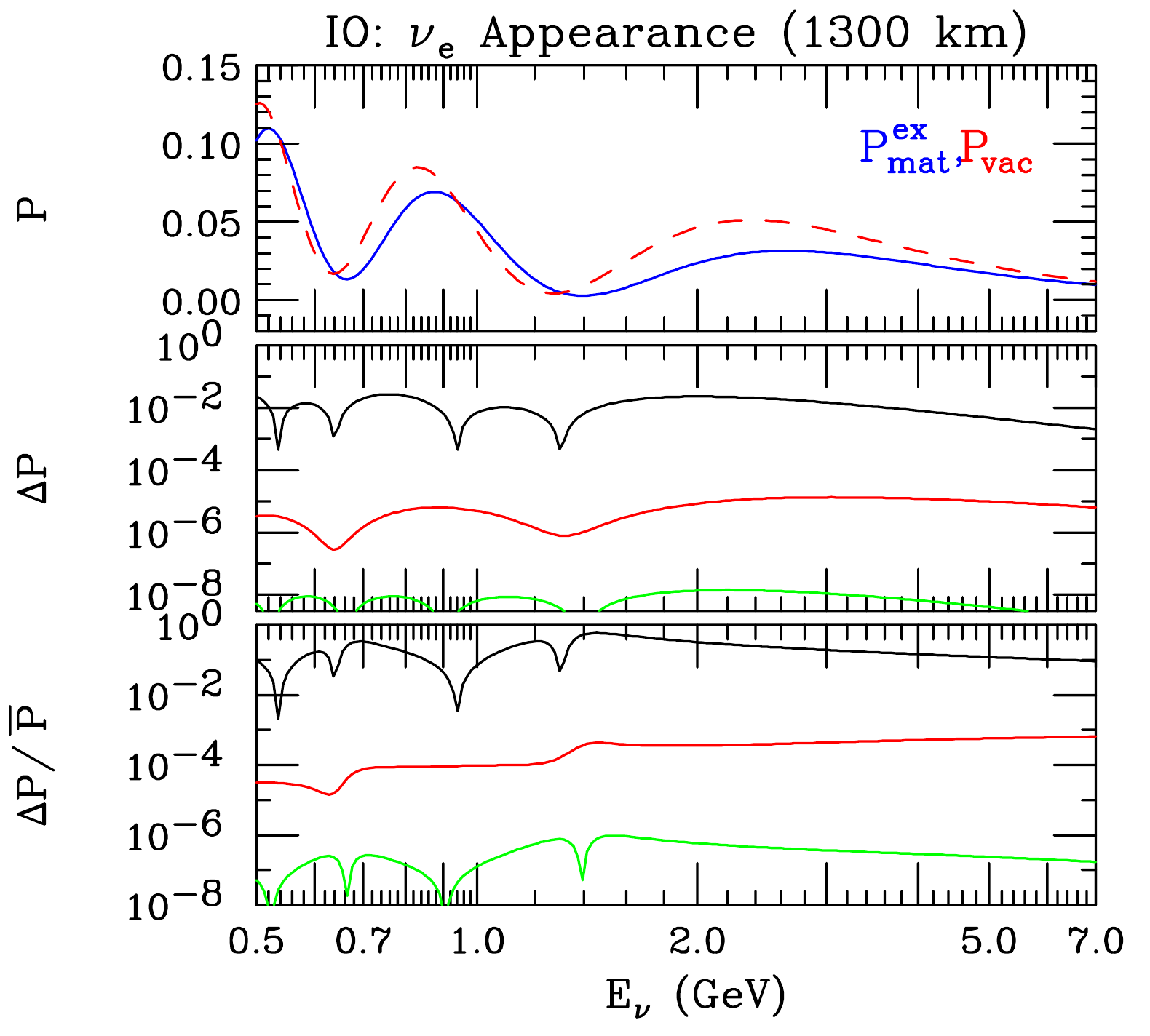}
     \includegraphics[width=.49\textwidth]{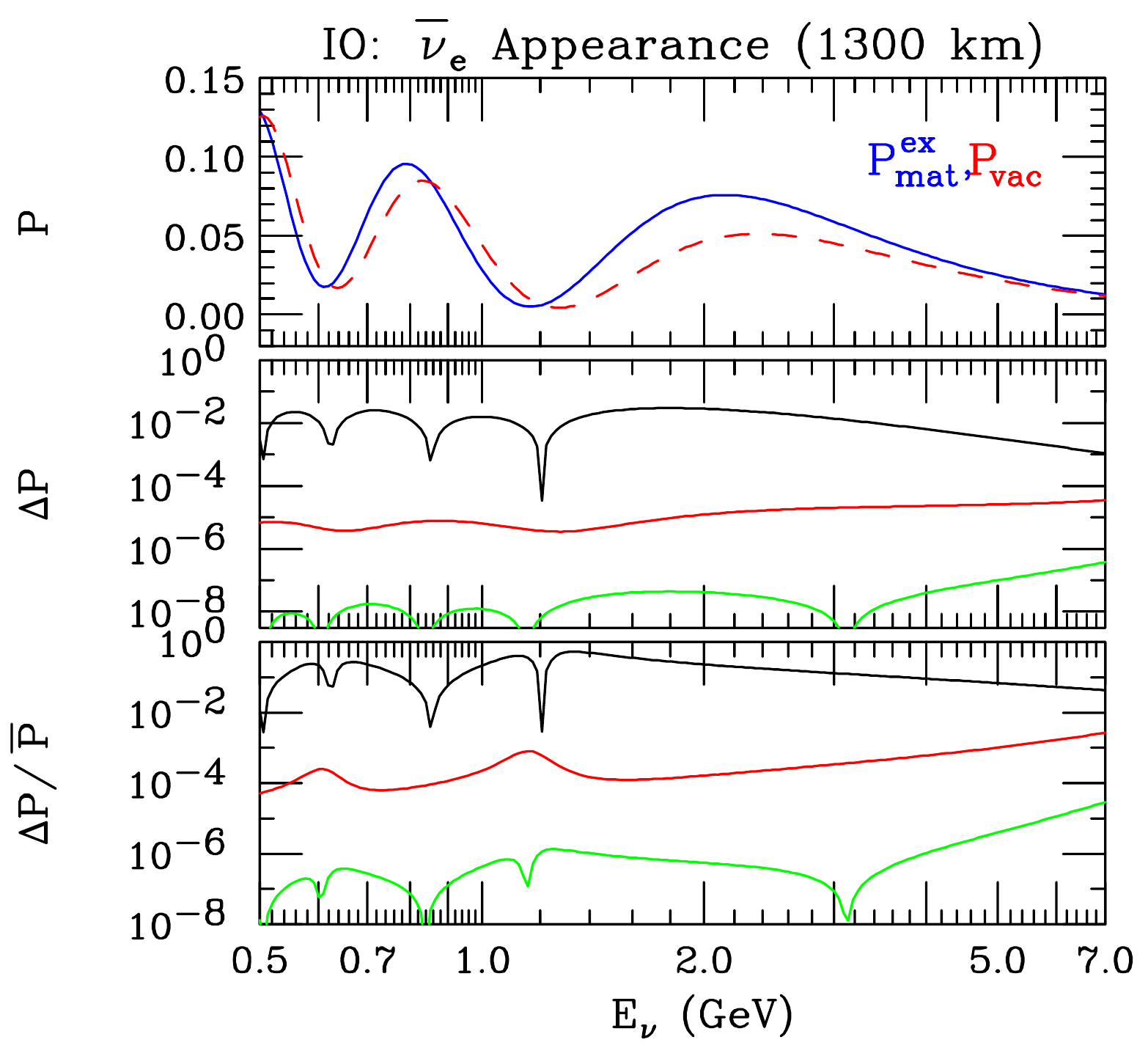}\\[5mm]
             \includegraphics[width=.95\textwidth]{DeltaP.pdf}\\
  \caption{DUNE, for inverted ordering (IO): Top Left figure is $\nu_\mu$ disappearance, Top Right figure is $\bar{\nu}_\mu$ disappearance, Bottom Left figure is $\nu_\mu \rightarrow \nu_e$ appearance, and Bottom Right is  $\bar{\nu}_\mu \rightarrow \bar{\nu}_e$ appearance.
  In each figure, the top panel is exact oscillation probability in matter , $P^{ex}_{mat}$,  from \cite{Zaglauer:1988gz}, and in vacuum, $P_{vac}$. The Middle panel is difference between exact oscillation probabilities in matter and vacuum (black), and the difference between exact and 0th (red) and exact and 1st (green) approximations to the matter probabilities  using the DMP scheme, \cite{Denton:2016wmg}. Bottom panel is similar to middle panel but plotting the fractional differences, $\Delta P/\overline{P}$. The density use is 3.0 g.cm$^{-3}$.
  }
     \label{fig:DUNE_IO}
          \end{center}

     \end{figure}

We consider four experimental setups: to be comprehensive, the energy windows are chosen to be wider than that accessible for a particular experiment.
\begin{itemize}
\item T2K \& T2HK:  with baseline, L = 295 km, neutrino energy 0.2 $ < E_\nu/{\rm GeV} <  $ 3.0, \\
and density, $\rho =$ 2.3 g.cm$^{-3}$. See  Fig. \ref{fig:T2K_NO}, \ref{fig:T2K_IO}.
\item NOvA:  with baseline, L = 810 km, neutrino energy 0.6 $ < E_\nu/{\rm GeV} <  $ 4.0, \\ and density,  $\rho =$ 3.0 g.cm$^{-3}$.
See Fig. \ref{fig:NOvA_NO}, \ref{fig:NOvA_IO}.
\item T2HKK:  with baseline, L = 1050 km, neutrino energy 0.3 $ < E_\nu/{\rm GeV} <  $ 5.0, \\and density,  $\rho =$ 3.0 g.cm$^{-3}$.
See  Fig. \ref{fig:T2HKK_NO}, Fig. \ref{fig:T2HKK_IO}.
\item DUNE:  with baseline, L = 1300 km, neutrino energy 0.5 $ < E_\nu/{\rm GeV} <  $ 7.0, \\ and density,  $\rho =$ 3.0 g.cm$^{-3}$.
See Fig. \ref{fig:DUNE_NO}, \ref{fig:DUNE_IO}.
\end{itemize}
In these figures we have considered the channels $\nu_\mu$ disappearance and $\nu_e$ appearance for neutrinos and anti-neutrinos and for both NO and IO.  Each figure consists of three panels: the top panel is the exact oscillation probabilities in matter from Zaglauer and Schwarzer, \cite{Zaglauer:1988gz}, as well as the exact vacuum oscillation probability. The middle (bottom) panel shows the difference (fractional difference) between\\[-5mm]
\begin{enumerate}
\item the exact, \cite{Zaglauer:1988gz}, and vacuum oscillation probabilities (black), 
\item the exact and the 0th order DMP approximation, \cite{Denton:2016wmg} (red), 
\item the exact and the 1st order DMP approximation, \cite{Denton:2016wmg}  (green). 
\end{enumerate}
In the fractional differences, the denominator is the average of the two probabilities being compared. In Fig, \ref{fig:T2K_NO} to Fig. \ref{fig:DUNE_IO}, the ``dips'' in middle and bottom panels appear when $\Delta P$ changes sign.

In Table \ref{tab:DP}, we give the maximum difference and fractional difference of the 0th order approximation to the exact probability.
\begin{table}[h]
\vspace*{-0.5cm}
\begin{center}
\begin{tabular}{|l|c|c|c|c|}
\hline
      &T2K/HK  & NOvA &T2HKK & DUNE \\
      \hline
max $\Delta P$ &  $10^{-5}$  &  $10^{-4}$ & $10^{-4}$ &  $10^{-4}$ \\
 max  $\Delta P/P $ & $10^{-3}$& $10^{-3}$ & $10^{-2.5}$ & $10^{-2}$ \\
 \hline
 \end{tabular}
\caption{The maximum $\Delta P$ and $\Delta P/P$ at 0th order in the DMP approximation. The largest fraction difference occurs at oscillation maximum for $\nu_\mu$ disappearance channel, where the oscillation probability is a few \%. None of the experiments included here, T2K, NOvA, T2HKK and DUNE will be within an order a magnitude  of being sensitive to any of these differences.}
\label{tab:DP}
\end{center}
\end{table}

\section{Summary}
In summary, the simple 0th order approximation of DMP,  \cite{Denton:2016wmg}, is sufficiently accurate for all of the accelerator based 
neutrino oscillation experiments operating or planned: T2K, NOvA, T2HKK and DUNE.  First order, which is also simple to use, improves the accuracy by a further two orders of magnitude for these experiments.

\section{Acknowledgements}

This manuscript has been authored by Fermi Research Alliance, LLC under Contract No. DE-AC02-07CH11359 with the U.S. Department of Energy, Office of Science, Office of High Energy Physics.

This project has received funding/support from the European Union's Horizon 2020 research and innovation programme under the Marie Sklodowska-Curie grant agreement No 690575.
 This project has received funding/support from the European Union's Horizon 2020 research and innovation programme under the Marie Sklodowska-Curie grant agreement No 674896.

HM is supported by Instituto F\'{\i}sica Te\'{o}rica, UAM/CSIC in Madrid, via ``Theoretical challenges of new high energy, astro and cosmo experimental data''  project, Ref: 201650E082. 

PBD acknowledges support from the Villum Foundation (Project No.~13164) and the Danish National Research Foundation (DNRF91 and Grant No.~1041811001).


\begin{thebibliography}{99}

\bibitem{Denton:2016wmg} 
  P.~B.~Denton, H.~Minakata and S.~J.~Parke,
  ``Compact Perturbative Expressions For Neutrino Oscillations in Matter,''
  JHEP {\bf 1606}, 051 (2016)
  doi:10.1007/JHEP06(2016)051
  [arXiv:1604.08167 [hep-ph]].

\bibitem{Parke:2018brr} 
  S.~J.~Parke, P.~B.~Denton and H.~Minakata,
  ``Analytic Neutrino Oscillation Probabilities in Matter: Revisited,''
  arXiv:1801.00752 [hep-ph].\\[2mm]
 S.J.~Parke and M.D.~Messier, ``Cross Check of NOvA Oscillation Probabilities''
  (1/12/2018), NOVA-doc-25833, FERMILAB-FN-1049,  http://doi.org/10.5281/zenodo.1146747 
  

\bibitem{Zaglauer:1988gz}
  H.~W.~Zaglauer and K.~H.~Schwarzer,
 ``The Mixing Angles in Matter for Three Generations of Neutrinos and the MSW Mechanism,''
  Z.\ Phys.\ C {\bf 40} (1988) 273.



\end{thebibliography}

\end{document}